\documentclass[12pt]{article}

\pdfoutput=1

\usepackage{graphicx}
\usepackage{epsfig}
\usepackage{amsfonts}
\usepackage{amssymb}
\usepackage{amsmath}
\usepackage{dsfont}
\usepackage{bbm}
\usepackage{multirow}
\usepackage[vcentermath]{youngtab}
\usepackage{latexsym}
\usepackage{verbatim}
\usepackage{mcite}
\usepackage{cite}
\usepackage{units}
\usepackage{cancel}
\usepackage[all]{xy}
\usepackage{eufrak}
\usepackage{tikz}

\usetikzlibrary{shapes,snakes,shadows,arrows}
\def\ellip{(0,0) ellipse (12 and 6)}
\def\ellipp{(2,0) ellipse (6 and 3)}
\tikzstyle{P_3} = [draw,very thick]
\tikzstyle{P_4} = [draw,fill=blue!20,very thick]

\def\beq{\begin{equation}}
\def\eeq{\end{equation}}
\def\beqa{\begin{eqnarray}}
\def\eeqa{\end{eqnarray}}

\def\bfone{\relax{\rm 1\kern-.35em 1}}

\newcommand{\be}{\begin{equation}}
\newcommand{\ee}{\end{equation}}
\newcommand{\ben}{\begin{displaymath}}
\newcommand{\een}{\end{displaymath}}
\newcommand{\bea}{\begin{eqnarray}}
\newcommand{\eea}{\end{eqnarray}}
\newcommand{\nn}{\nonumber}

\newcommand{\bean}{\begin{eqnarray*}}
\newcommand{\eean}{\end{eqnarray*}}

\begin{document}
\pagestyle{plain}


\makeatletter \@addtoreset{equation}{section} \makeatother
\renewcommand{\thesection}{\arabic{section}}
\renewcommand{\theequation}{\thesection.\arabic{equation}}
\renewcommand{\thefootnote}{\arabic{footnote}}


\setcounter{page}{1} \setcounter{footnote}{0}


 \begin{titlepage}
 \begin{flushright}
 \small ~~
 \end{flushright}

 \bigskip

 \begin{center}

 \vskip 0cm

 {\LARGE \bf Duality orbits of non-geometric fluxes}\\[6mm]
 \vskip 0.5cm

 {\bf G. Dibitetto$^{1}$,\, J. J. Fern\'andez-Melgarejo$^{2}$,\, D. Marqu\'es$^{3}$ \,and\, D. Roest$^{1}$}\\

\vskip 25pt

 {\em $^1$ Centre for Theoretical Physics,\\
 University of Groningen, \\
 Nijenborgh 4, 9747 AG Groningen, The Netherlands\\
 {\small {\tt \{ g.dibitetto, d.roest \} @rug.nl}}} \\

\vskip 0.3cm

 {\em $^2$ Grupo de F\'isica Te\'orica y Cosmolog\'ia, Dept. de F\'isica, \\
 University of Murcia,\\
 Campus de Espinardo, E-30100-Murcia, Spain\\
 {\small {\tt jj.fernandezmelgarejo@um.es}}}

 \vskip 0.3cm

 {\em $^3$ Institut de Physique Th\'eorique, \\
 CEA/ Saclay , 91191 Gif-sur-Yvette Cedex, France\\
 {\small {\tt diego.marques@cea.fr}}}

\vskip 0.8cm

 \end{center}

 \vskip 1cm

 \begin{center}

 {\bf ABSTRACT}\\[3ex]

 \begin{minipage}{13cm}
 \small

Compactifications in duality covariant constructions such as
generalized geometry and double field theory have proven to be suitable
frameworks to reproduce gauged supergravities containing
non-geometric fluxes. However, it is a priori unclear whether these
approaches only provide a reformulation of old results, or also
contain new physics. To address this question, we classify the T-
and U-duality orbits of gaugings of (half-)maximal supergravities in
dimensions seven and higher. It turns out that all orbits have a
geometric supergravity origin in the maximal case, while there are
non-geometric orbits in the half-maximal case. We show how the
latter are obtained from compactifications of double field theory.

\end{minipage}

\end{center}

\vfill

\end{titlepage}


\tableofcontents

\section{Introduction}
\label{sec:introduction}

When compactifying heterotic, type II or eleven-dimensional supergravity on a
given background, one obtains lower-dimensional effective theories
whose features depend on the fluxes included in the compactification
procedure and, in particular, on the amount of supersymmetry
preserved by the chosen background. When some supersymmetry is
preserved during the compactification, the effective theories under consideration
are then gauged supergravities.

In particular, in the context of half-maximal  \cite{Schon:2006kz}
and maximal \cite{ deWit:2007mt} gauged supergravities, not only
does supersymmetry tightly organise the ungauged theory, but also it
strictly determines the set of possible deformations (\emph{i.e.}
gaugings). The development of the so-called embedding tensor
formalism has enabled one to formally describe all the possible
deformations in a single universal formulation, which therefore
completely restores duality covariance. Unfortunately, not all the
deformations have a clear higher-dimensional origin, in the sense
that they can be obtained by means of a certain compactification of
ten or eleven dimensional supergravity.

One of the most interesting open problems concerning flux
compactifications is to reproduce, by means of a suitable flux
configuration, a given lower-dimensional gauged supergravity theory.
Although this was done in particular cases (see for example
\cite{Roest:2009dq, Dall'Agata:2009gv}), an exhaustive analysis
remains to be done. This is due to fact that, on the one hand we
lack a classification of the possible gauging configurations allowed
in gauged supergravities and, on the other hand, only a limited set
of compactification scenarios are known. Typically, to go beyond the
simplest setups one appeals to dualities. The paradigmatic example
\cite{Shelton:2005cf} starts by applying T-dualities to a simple
toroidal background with a non-trivial two-form generating a single
$H_{abc}$ flux. By T-dualizing this setup, one can construct a chain
of T-dualities leading to new backgrounds (like twisted-tori or
T-folds) and generating new (dual) fluxes, like the so-called
${Q_a}^{bc}$ and $R^{abc}$. It is precisely by following duality
covariance arguments in the lower-dimensional effective description
that non-geometric fluxes \cite{Shelton:2005cf} were first
introduced in order to explain the mismatch between particular flux
compactifications and generic gauged supergravities.

\begin{figure}[h!]
\begin{center}
\begin{tikzpicture}[scale=0.5,>=latex']
        \path[P_3] \ellip;
        \path[P_4] \ellipp;
        \draw[very thick] (-9,4) -- (-5,-5.45);
        \draw[very thick] (-4,5.65) -- (1,-6);
        \path (-3,2.25) node[left] {\textbf{A}}
            (-1.5,-3.5) node [right] {\textbf{B}};
        \path (-8,0) node[left] {orbit 1}
            (1,-4.5) node [right] {orbit 2};
        \path (16.5,-5) node[left] {\textbf{Flux configurations}}
            (0,.5) node [right] {\textbf{Geometric}};
        \path (-.5,-.75) node[right] {\textbf{configurations}};
\end{tikzpicture}
{\it \caption{\it The space of flux configurations sliced into
duality orbits (vertical lines). Moving along a given orbit
corresponds to applying dualities to a certain flux configuration
and hence it does not imply any physical changes in the
lower-dimensional effective description. Geometric fluxes only
constitute a subset of the full configuration space. Given an orbit,
the physically relevant question is whether (orbit 2 between A and
B) or not (orbit 1) this intersects the geometric subspace. We refer
to a given point in an orbit as a {\it
representative}.}\label{pic:orbits}}
\end{center}
\end{figure}
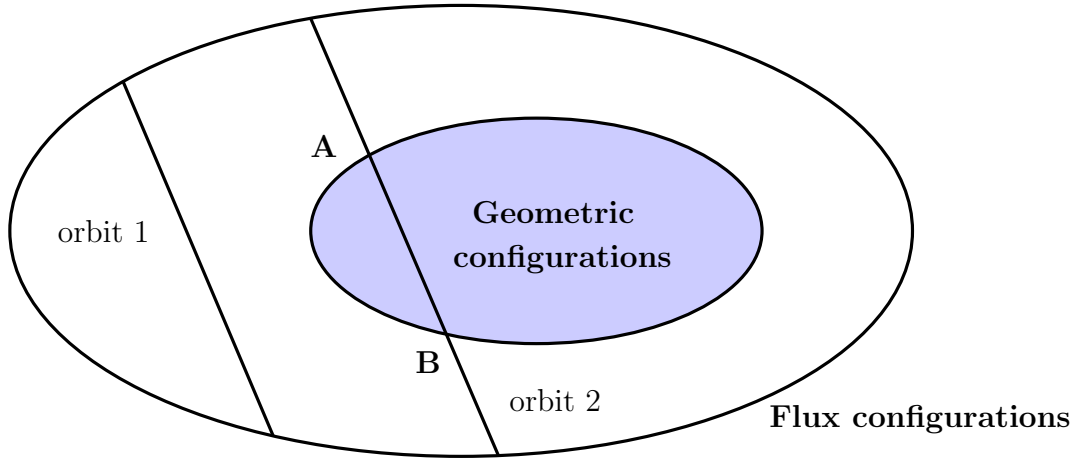

Here we would like to emphasize that all these (a priori) different
T-duality connected flux configurations by definition lie in the
same orbit of gaugings,  and therefore give rise to the same
lower-dimensional physics. In order to obtain a different gauged
supergavity, one should consider more general configurations of
fluxes, involving for example combinations of geometric and
non-geometric fluxes, that can never be T-dualised to a frame in
which the non-geometric fluxes vanish. For the sake of clarity, we
depict this concept in figure~\ref{pic:orbits}.

Non-geometric fluxes are
the inevitable consequence of string dualities, and only a theory
which promotes such dualities to symmetries could have a
chance to describe them together with geometric fluxes and to understand their origin in a unified way. From the
viewpoint of the lower-dimensional effective theory, it turns out
that half-maximal and maximal gauged supergravities give
descriptions which are explicitly covariant with respect to T- and
U-duality respectively. This is schematically depicted in table~
\ref{dualities}, even though only restricted to the cases we will address in this work.

\begin{table}[h]
\renewcommand{\arraystretch}{1.25}
\begin{center}
\scalebox{0.85}[0.85]{
\begin{tabular}{|c|c|c|}
\hline
$D$ & T-duality & U-duality \\
\hline \hline
$9$ &  O($1,1$) & $\mathbb{R}^{+}\,\times\,$SL($2$)\\
\hline
$8$ &  O($2,2)\,=\,$SL($2)\,\times\,$SL($2$) & SL($2)\,\times\,$SL($3$)\\
\hline
$7$ &  O($3,3)\,=\,$SL($4$) & SL($5$)\\
\hline
\end{tabular}
}
\end{center}
{\it \caption{The various T- and U-duality groups in $D>6$. These turn
out to coincide with the global symmetry groups of half-maximal and
maximal supergravities respectively.} \label{dualities}}
\end{table}

In recent years, a new proposal aiming to promote T-duality to a
fundamental symmetry in field theory has received increasing interest. It is named Double Field Theory (DFT)
\cite{Hull:2009mi} since T-duality invariance requires
a doubling of the spacetime coordinates, by supplementing them with
dual coordinates associated to the stringy winding modes, whose dynamics can
become important in the compactified theory. Recently it has been
pointed out how to obtain gaugings of $\mathcal{N}=D=4$ supergravity
by means of twisted double torus reductions of DFT
\cite{Aldazabal:2011nj, Geissbuhler:2011mx}, even though at that
stage, the so-called \emph{weak} and \emph{strong} constraints imposed for
consistency of DFT represented a further restriction that prevented
one from describing the most general gaugings that solve the Quadratic
Constraints (QC) of gauged supergravity.

Subsequently, an indication has been
given that gauge consistency of DFT does not need the weak and strong
constraints \cite{Grana:2012rr}. Following this direction, we could wonder whether
relaxing these constraints can provide a higher-dimensional origin
for all gaugings of extended supergravity in DFT. Our aim in the present
work is to assess to what extent DFT can improve our description of
non-geometric fluxes by giving a higher-dimensional origin to orbits
which do not follow from standard supergravity compactifications. We will
call such orbits of gaugings {\it non-geometric} (in figure \ref{pic:orbits} they are represented by orbit 1).

As a starting point for this investigation, we will address the
problem in the context of maximal and half-maximal gauged
supergravities in seven dimensions and higher, where the global
symmetry groups are  small enough to allow for a general
classification of orbits, without needing to consider truncated
sectors. We will show that in the half-maximal supergravities in
seven and higher-dimensions, where the classifications of orbits can
be done exhaustively, {\it all} the orbits (including geometric and
non-geometric) admit an uplift to DFT, through Scherk-Schwarz (SS)
\cite{Scherk:1979zr} compactifications on appropriate backgrounds.
We provide explicit backgrounds for every orbit, and discuss their
(un)doubled nature. The result is that truly doubled DFT provides the appropriate framework to deal with orbits that can not be obtained from supergravity. In contrast, in maximal supergravities in eight and higher-dimensions, all
orbits are geometric and hence can be obtained without resorting to DFT.

The paper is organised as follows.  In section~\ref{sec:DFT} we
start with a brief review and motivation of DFT. We will make
particular emphasis in discussing aspects of its SS
compactifications, and the constraints arising from gauge
consistency. We will explicitly show how the gaugings in the
effective theory are related to the compactification ansatz, in
order to make a link with the results of the following sections. In
section~\ref{sec:U_Dualitites} we present the classification of
consistent gaugings in maximal supergravity in terms of U-duality
orbits. In particular, in section~\ref{subsec:Max9}
and~\ref{subsec:Max8}, we work out the $D=9$ and $D=8$ orbits. In
both cases we are able to show that all the duality orbits have a
geometric origin in compactifications of ten dimensional
supergravity. In section~\ref{sec:T_Dualitites} we classify the
consistent gaugings in half-maximal supergravity in terms of
T-duality orbits. In particular, in section~\ref{subsec:Half_Max8}
and~\ref{subsec:Half_Max7}, we work out the $D=8$ and $D=7$ orbits.
Here we encounter the first orbits lacking a geometric
higher-dimensional origin. We show that such orbits do follow from
dimensional reductions of DFT. Finally, our conclusions are
presented in section~\ref{sec:conclusions}. We defer a number of technical
details on gauge algebras and 't Hooft symbols to the appendices.

\section{Orbits from double field theory}
\label{sec:DFT}

When two configurations of gaugings are connected by a duality
transformation, the physics they give rise to is the same. In this
direction, we have defined an {\it orbit} of gaugings as a set of
gauged theories that are related by dualities. One can then state
that physically distinct theories are labeled by orbits, rather than
by generic solutions to the QC. In this section we will provide the
link between orbits in gauged supergravities and SS
compactifications of DFT.

DFT is a recent proposal that promotes T-duality to a symmetry in
field theory \cite{Hull:2009mi, Hull:2009zb}, and is currently
defined in terms of a background independent action
\cite{Hohm:2010jy, *Hohm:2010pp}. The theory is defined on a double
space \cite{Hull:2004in, *Dabholkar:2005ve}, and its original
version was created to describe the dynamics of closed strings on
tori, the dual coordinates being associated to the winding modes of
the strings. However, the background independent action allows for
more general spaces, and SS compactifications of DFT were shown to
formally reproduce the bosonic (electric) sector of half-maximal
gauged supergravities \cite{Aldazabal:2011nj, Geissbuhler:2011mx}.
The gauge invariance of DFT and closure of its gauge algebra gives
rise to a set of constraints that restrict the coordinate dependence
of the fields. A possible set of solutions to such constraints is
given by restricting the fields and gauge parameters to satisfy the so-called weak and strong constraints.
In such a situation, they can always be T-dualised to a frame in which the
dependence on dual coordinates is cancelled. This restriction arises
naturally in the context of toroidal compactifications and has a
close relation to the level-matching condition in the sigma model.
In such case, DFT provides an interesting framework in which
ten-dimensional supergravity can be rotated to T-dual frames
\cite{Andriot:2011uh, *Andriot:2012wx}. Many other interesting works
on the subject can be found in \cite{Hohm:2011ex,
 *Hohm:2010xe, Jeon:2010rw, *Berman:2010is, *Berman:2011jh, *Berman:2011pe, *Coimbra:2011ky, *Coimbra:2011nw, *West:2011mm , *Hohm:2011nu, *Jeon:2011sq,
*Copland:2011wx, *Berman:2011kg, *Jeon:2011vx, *Jeon:2011cn, *Kan:2011vg,
*Hohm:2011dv, *Albertsson:2011ux, *Thompson:2011uw,
 *Jeon:2011kp,
*Hohm:2011si, *Vaisman:2012ke}.

While toroidal compactifications of DFT lead to half-maximal
ungauged supergravities, SS compactifications on more general double
spaces are effectively described by gauged supergravities like the
ones we will analyse in the next sections. If the internal space is
restricted in such a way that there always exists a frame without
dual coordinate dependence, the only orbits allowed in the effective
theory are those admitting representatives that can be obtained from
 compactifications of ten dimensional supergravity. This is not the most general case, and we will show that some orbits require
the compact space to be truly doubled, capturing information of both momentum and winding modes.

Recently in ref.~\cite{Grana:2012rr}, a new set of solutions to the
constraints for DFT has been found. For these solutions the internal
dependence of the fields is not dynamical, but fixed. The
constraints of DFT restrict the dynamical external space to be
undoubled, but allows for a doubling of the internal coordinates as
long as the QC for the gaugings are satisfied. Interestingly, these
are exactly the constraints needed for consistency of gauged
supergravity, so there is a priori no impediment to uplift any orbit
to DFT in this situation. In fact, in the following sections we show
that all the orbits in half-maximal $D = 7,8$ gauged supergravities
can be reached from twisted double tori compactifications of DFT.

\subsection{DFT and (half-)maximal gauged supergravities}
\label{subsec:GDFT}

DFT is a field theory with manifest invariance under the O$(d,d)$
T-duality group, and therefore captures stringy features. The coordinates form fundamental vectors $X^M =
(\tilde x_i, x^i)$, containing $d$ space-time coordinates $x^i$ and
$d$ dual coordinates $\tilde x_i$, $i=1,...,d$. The field content is
that of the NS-NS sector, but defined on the double space. The
generalised metric is a symmetric element of O$(d,d)$ \be {\cal
H}_{MN} = \begin{pmatrix} g^{ij} & -g^{ik} b_{kj} \\ b_{ik} g^{kj} &
g_{ij} - b_{ik} g^{kl} b_{lj}\end{pmatrix} \ee and includes the
$d$-dimensional metric $g_{ij}$, and  the $d$-dimensional
Kalb-Rammond field $b_{ij}$. The metric of the global symmetry group
 \be
 \eta_{MN} = \begin{pmatrix}0^{ij} & \delta^i{}_j \\ \delta_i{}^j & 0_{ij}\end{pmatrix} \label{lightcone}
 \ee
raises and lowers the indices of $\cal H$, such that ${\cal H}_{MP}
{\cal H}^{PN} = \delta^N_M$. On the other hand, the dilaton $\phi$
is combined with the determinant of $g$ in a $T$-invariant way
$e^{-2d} = \sqrt{g} e^{-2\phi}$.

Detailed  reviews of DFT can be found in refs~\cite{Hohm:2011gs,
*Zwiebach:2011rg}. Here we will only provide a discussion of its
constraints and some aspects of its SS reductions, just the minimal
ingredients with the corresponding references to make contact with
the results of the following sections.

In the SS procedure, the coordinates $X^M$ are split into  external directions $\mathbb{X}$ and compact internal
$\mathbb{Y}$ coordinates.  The former set contains pairs of O$(D,D)$
dual coordinates, while the latter one contains pairs of O$(n,n)$ dual
coordinates, with $d = D + n$. This means that if a given coordinate
is external (internal), its dual must also be external (internal),
so the effective theory is formally a (gauged) DFT. The SS procedure
is then defined in terms of a reduction ansatz, that specifies the
dependence of the fields in $(\mathbb{X},\mathbb{Y})$ \be {\cal
H}_{MN} (\mathbb{X},\mathbb{Y}) = U(\mathbb{Y})^A{}_M \ \widehat
{\cal H}(\mathbb{X})_{AB}\ U(\mathbb{Y})^B{}_N
 \ , \ \ \ \ \ d(\mathbb{X},\mathbb{Y}) = \widehat d(\mathbb{X}) + \lambda(\mathbb{Y})\ .\ee
Here the hatted fields $\widehat {\cal H}$ and $\widehat{d}$  are
the dynamical fields in the effective theory, parameterizing
perturbations around the background, which is defined by
$U(\mathbb{Y})$ and $\lambda(\mathbb{Y})$. The matrix $U$ is
referred to as the \emph{twist matrix}, and must be an element of
O$(n,n)$. It contains a DFT T-duality index $M$, and another index
$A$ corresponding to the T-duality group of the effective theory.
When DFT is evaluated on the reduction ansatz, the twists generate
the gaugings of the effective theory \bea f_{ABC} &=& 3 \eta_{D[A}\
(U^{-1})^M{}_B (U^{-1})^N{}_{C]} \partial_M U^D{}_M \
,\label{f_from_U_}
\\
\xi_A &=& \partial_M (U^{-1})^M{}_A - 2 (U^{-1})^M{}_A \partial_M
\lambda \ ,\label{f_from_U}\eea
where $f_{ABC}$ and $\xi_{A}$ build the generalised structure
constants of the gauge group in the lower-dimensional theory.

Although $U$ and $\lambda$ are $\mathbb{Y}$ dependent quantities,
the gaugings are forced to be constants in order to eliminate the
$\mathbb{Y}$ dependence from the lower dimensional theory.
 When the external-internal splitting is performed, namely $d = D + n$, the dynamical fields are written in terms of their components which are a $D$-dimensional metric, a $D$-dimensional $2$-form, $2n$ $D$-dimensional vectors and $n^2$ scalars. These are the degrees of freedom of half-maximal supergravities. Since these fields are contracted with the gaugings, one must make sure that after the splitting the gaugings have vanishing Lorentzian indices, and this is achieved by stating that the twist matrix is only non-trivial in the internal directions. Therefore, although formally everything is covariantly written in terms of O$(d,d)$ indices $A,B,C,...$, the global symmetry group is actually broken to O$(n,n)$. We will not explicitly show how this splitting takes place, and refer to \cite{Aldazabal:2011nj} for more details. In this work,  for the sake of simplicity, we will restrict to the case $\xi_A  =0$, which should be viewed as a constraint for $\lambda$. Also we will restrict to $O(n,n)$ global symmetry groups, without additional vector fields.

There are two possible known ways to restrict the fields and gauge
parameters in DFT, such that the action is gauge invariant and the
gauge algebra closes. On the one hand, the so-called weak and strong
constraints can be imposed \be
\partial_M \partial^M A = 0\ , \ \ \ \ \partial_M A\ \partial^M B =
0\ ,\ee where $A$ and $B$ generically denote products  of
(derivatives of) fields and gauge parameters. When this is the case,
one can argue \cite{Hohm:2010jy} that there is always a frame in
which the fields do not depend on the dual coordinates. On the other
hand, in the SS compactification scenario, it is enough to impose
the weak and strong constraints only on the external space
(\emph{i.e.}, on hatted quantities) \be
\partial_M \partial^M \widehat A = 0\ , \ \ \ \ \partial_M \widehat A\ \partial^M \widehat B = 0
\ ,\ee and impose QC for the gaugings \be f_{E[AB} f^E{}_{C]D}
= 0 \ .\ee This second option is more natural for our purposes,
since these constraints exactly coincide with those of half-maximal
gauged supergravities\footnote{We are working under the assumption that the structure constants not only specify the gauging, but all couplings of the theory. Reproducing the correct structure constants therefore implies reproducing the full theory correctly, as has been proven in $D=4$ and $D=10$ \cite{Aldazabal:2011nj, Geissbuhler:2011mx, Hohm:2011ex,
 *Hohm:2010xe}.} (which are undoubled theories in the external
space, and contain gaugings satisfying the QC).

Notice that if a given $U$ produces a solution to the QC, any T-dual
$U$ will also. Therefore, it is natural to define the notion of {\it
twist orbits}  as the sets of twist matrices connected through
T-duality transformations. If a representative of a twist orbit
generates a representative of an orbit of gaugings, one can claim
that the twist orbit will generate the entire orbit of gaugings.
Also, notice that if a twist matrix satisfies the weak and strong
constraints, any representative of its orbit will, so one can define
the notions of undoubled and truly doubled twist orbits.

\subsubsection*{Non-geometry VS weak and strong constraint violation}

Any half-maximal supergravity can be uplifted to the maximal theory
whenever the following constraint holds\footnote{$D=4$ half-maximal
supergravity is slightly different because its global symmetry group
features an extra SL($2$) factor; for full details, see \cite{Aldazabal:2011yz, Dibitetto:2011eu}.}
\be f_{ABC}\,f^{ABC}\,=\,0\ . \label{Extra_f}\ee
This constraint plays the role of an orthogonality condition between geometric and non-geometric fluxes.
Interestingly, the constraint \eqref{Extra_f} evaluated in terms of
the twist matrix $U$ and $\lambda$ can be rewritten as follows (by taking
relations \eqref{f_from_U_} and \eqref{f_from_U} into account)
\be
f_{ABC}\,f^{ABC}\,=\,-3\,\partial_{D}{U^{A}}_{P}\,\partial^{D}{\left(U^{-1}\right)^{P}}_{A}-24\,\partial_{D}\lambda\,\partial^{D}\lambda\,+\,24\,\partial_{D}\partial^{D}\lambda\
. \label{Max_VS_Geom}\ee
The RHS of this equation is zero whenever the background defined by
$U$ and $\lambda$ satisfies the weak and strong constraints. This immediately
implies that any background satisfying weak and strong constraints
defines a gauging which is upliftable to the maximal theory.
Conversely, if an orbit of gaugings in half-maximal supergravity
does not satisfy the extra constraint \eqref{Extra_f}, the RHS of
this equation must be non-vanishing, and then the strong and weak
constraint must be relaxed. In conclusion, the orbits of
half-maximal supergravity that do not obey the QC of the maximal
theory require truly doubled twist orbits, and are therefore
genuinely non-geometric. This point provides a concrete criterion to
label these orbits as non-geometric. Also, notice that these orbits
will never be captured by non-geometric flux configurations obtained
by T-dualizing a geometric background\footnote{However, we would like to
stress that, in general, it is not true that an orbit satisfying the
QC constraints of maximal supergravity (\ref{Extra_f}) is
necessarily generated by an undoubled twist orbit. An example can be found at the end of section 4.}.

For the sake of clarity, let us briefly review the definitions that we use. A twist orbit is non-geometric if it doesn't satisfy the weak/strong constraint, and geometric if it does. Therefore, the notion of geometry that we consider is local, and we will not worry about global issues (given that the twist matrix is taken to be an element of the global symmetry group, the transition functions between coordinate patches are automatically elements of $O(n,n)$). On the other hand an orbit of gaugings is geometric if it contains a representative that can be obtained from 10 dimensional supergravity (or equivalently from a geometric twist orbit), and it is non-geometric if it does not satisfy the constraints of maximal supergravity.

We have now described all the necessary ingredients to formally relate dimensional reductions of DFT and the orbits
of half-maximal gauged supergravities. In particular, in what follows we will:
\begin{enumerate}
\item Provide a classification of all the orbits of gaugings in maximal and half-maximal supergravities in $D\geq7$.
\item Explore mechanisms to generate orbits of gaugings from twists, satisfying
\begin{itemize}
\item $U(\mathbb{Y}) \in \textrm{O}(n,n)$
\item Constant $f_{ABC}$
\item  $f_{E[AB} f^E{}_{C]D} = 0$
\end{itemize}
\item Show that in the half-maximal theories all the orbits of gaugings  can be obtained from twist orbits in DFT.
\item Show that in the half-maximal theories the orbits that satisfy the QC of maximal supergravity  admit a representative with a higher-dimensional supergravity origin. For these we provide concrete realisations in terms of unboubled backgrounds in DFT. Instead, the orbits that fail to satisfy (\ref{Extra_f}) require, as we argued, truly doubled twist orbits for which we also provide concrete examples.
\item Show that  there is a degeneracy in the space of twist orbits giving rise to the same orbit of gaugings. Interestingly, in some cases a given orbit can be obtained either from undoubled or truly doubled twist orbits.
\end{enumerate}

In the next sections we will classify all the orbits in (half-)maximal $D\geq 7$ supergravities, and provide the half-maximal ones with concrete uplifts to DFT, explicitly proving the above points.

\subsection{Parametrisations of the duality twists}
\label{subsec:twist_matrices}

Here we would like to introduce some notation that will turn out to
be useful in the uplift of orbits to DFT. We start by noting the
double internal  coordinates as $\mathbb{Y}^A = (\tilde y_a, y^a)$
with $a = 1,...,n$. As we saw, the SS compactification of DFT is
defined by the twists $U(\mathbb{Y})$ and $\lambda(\mathbb{Y})$. The
duality twist $U(\mathbb{Y})$ is not generic, but forced to be an
element of O($n,n$), so we should provide suitable
parameterisations. One option is the  {\it light-cone}
parameterisation, where the metric of the (internal) global symmetry
group is taken to be of the form (\ref{lightcone}) \be \eta_{AB} =
\begin{pmatrix} 0 & \mathds{1}_n \\ \mathds{1}_n & 0\end{pmatrix}\ . \ee The most
general form of the twist matrix is then given by \be U(\mathbb{Y}) =
\left(\begin{matrix} e & 0 \\ 0 & e^{-T}\end{matrix}\right)
\,\left(\begin{matrix} \mathds{1}_n & 0 \\ -B &
\mathds{1}_n\end{matrix}\right)\, \left(\begin{matrix} \mathds{1}_n & \beta \\
0 & \mathds{1}_n\end{matrix}\right)\ , \ee with $e \in
\textrm{GL}(n)$ and $B$ and $\beta$ are generic $n\times n$
antisymmetric matrices. When $\beta = 0$, $e = e(y^a)$ and $B =
B(y^a)$, the matrix $e$ can be interpreted as a $n$-dimensional
internal vielbein and $B$ as a background $2$-form for the
$n$-dimensional internal Kalb-Ramond field $b$. Whenever the
background is of this form, we will refer to it as geometric (notice
that this still does not determine completely the background, which
receives deformations from scalar fluctuations). In this case the
gaugings take the simple form \bea
f_{abc} &=& 3 (e^{-1})^\alpha{}_{[a}(e^{-1})^\beta{}_{b}(e^{-1})^\gamma{}_{c]} \partial_{[\alpha} B_{\beta\gamma]}\ , \nn\\
f^a{}_{bc} &=& 2 (e^{-1})^\beta{}_{[b}(e^{-1})^\gamma{}_{c]}\partial_{\beta} e^a{}_{\gamma }\ , \nn\\
f^{ab}{}_c &=& f^{abc} = 0 \ .\eea

If we also turn on a  $\beta (y^a)$, the relation of $e$, $B$ and
$\beta$ with the internal $g$ and $b$ is less trivial, and typically
the background will be globally well defined up to O$(n,n)$
transformations mixing the metric and the two-form (this is typically called a T-fold). In this case,
we refer to the background as locally geometric but globally
non-geometric, and this situation formally allows for non-vanishing
$f^{ab}{}_c$ and $f^{abc}$. Finally, if the twist matrix is a
function of $\tilde y_a$, we refer to the background as locally
non-geometric. Notice however, that if it satisfies the weak and
strong constraints, one would always be able to rotate it to a frame
in which it is locally geometric, and would therefore belong to an
undoubled orbit.

Alternatively, one could also define the  {\it cartesian}
parametrisation of the twist matrix, by taking
 the metric of the (internal) global symmetry group to be of the form
\be \eta_{AB} = \begin{pmatrix} \mathds{1}_n & 0 \\ 0 &
-\mathds{1}_n\end{pmatrix} \ .\ee This formulation is related to the
light-cone parametrisation through a SO$(2n)$ transformation, that
must also rotate the coordinates. In this case the relation between
the components of the twist matrix and the internal $g$ and $b$ is
non-trivial. We will consider the O($n,n$) twist
matrix to contain a smaller O($n-1,n-1$) matrix in the directions
$(y^2,...,y^n,\tilde y_2,...,\tilde y_n)$ fibred over the flat
directions $(y^1,\tilde y_1)$. We have seen that this typically
leads to constant gaugings.

Of course these are not the most general parameterisations and
ansatz, but they will serve our purposes of uplifting all the orbits
of half-maximal supergravity to DFT. Interesting works on how to
generate gaugings from twists are \cite{Dall'Agata:2007sr,
*Andriot:2009fp}.

\section{U-duality orbits of maximal supergravities}
\label{sec:U_Dualitites}

Following the previous discussion of DFT and its relevance for generating duality orbits, we turn to the actual classification of these. In particular, we start with orbits under U-duality of gaugings of maximal supergravity. Moreover, we will demonstrate that all such orbits do have a higher-dimensional supergravity origin.

Starting with the highest dimension for  maximal supergravity,
$D=11$, no known deformation is possible here. Moreover, in $D=10$
maximal supergravities, the only possible deformation occurs in what
is known as massive IIA supergravity\footnote{Throughout this paper
we will not consider the trombone gaugings giving rise to theories
without an action principle, as discussed in
\emph{e.g.}~\cite{Howe:1997qt, Bergshoeff:2002nv, LeDiffon:2008sh,
LeDiffon:2011wt}.} \cite{Romans:1985tz}. It consists of a
St\"uckelberg-like way of giving a mass to the 2-form $B_{2}$.
Therefore, such a deformation cannot be interpreted as a gauging.
The string theory origin of this so-called Romans' mass parameter is
nowadays well understood as arising from D8-branes
\cite{Polchinski:1995mt}. Furthermore, its DFT uplift has been
constructed in ref.~\cite{Hohm:2011cp}. Naturally, the structure of
possible orbits becomes richer when going to lower dimensions. In what follows we will perform
the explicit classification in dimensions nine and eight.

\subsection{Orbits and origin of the $D=9$ maximal case}
\label{subsec:Max9}

\subsubsection*{Maximal $D=9$ gauged supergravity}

The maximal (ungauged) supergravity in $D=9$ \cite{Gates:1984kr} can
be obtained by reducing either massless type IIA or type IIB
supergravity in ten dimensions on a circle. The global symmetry
group of this theory is
\be G_{0}\,=\,\mathbb{R}^{+}\,\times\,\textrm{SL}(2)\ . \notag\ee
Note that $G_{0}$ is the global symmetry of the action and hence it
is realised off-shell, whereas the on-shell symmetry has an extra
$\mathbb{R}^{+}$ with respect to which the Lagrangian has a
non-trivial scaling weight. This is normally referred to as the
\emph{trombone symmetry}. As a consequence, the on-shell symmetry
contains three independent rescalings \cite{Bergshoeff:2002nv,
Roest:2004pk}, which we summarise in table~\ref{rescalings}.
\begin{table}[h!]
\begin{center}
\scalebox{1}[1]{
\begin{tabular}{| c || c | c | c | c | c | c | c | c | c | c | c | c| c|}
\hline
\textrm{ID} & $e_{\mu}^{\phantom{\mu}a}$ & $A_{\mu}$ &  $A_{\mu}{}^{1}$ & $A_{\mu}{}^{2}$ & $B_{\mu\nu}{}^{1}$ & $B_{\mu\nu}{}^{2}$ & $C_{\mu\nu\rho}$ & $e^{\varphi}$ & $\chi$ & $e^{\phi}$ & $\psi_{\mu}$ & $\lambda\,,\,\tilde{\lambda}$ & $\mathcal{L}$  \\[1mm]
\hline \hline
$\alpha$ & $\frac{9}{7}$ & $3$ & $0$ & $0$ & $3$ & $3$ & $3$ & $\frac{6}{\sqrt{7}}$ & $0$ & $0$ & $\frac{9}{14}$ & $-\frac{9}{14}$ & $9$ \\[1mm]
\hline
$\beta$ & $0$ & $\frac{1}{2}$ & $-\frac{3}{4}$ & $0$ & $-\frac{1}{4}$ & $\frac{1}{2}$ & $-\frac{1}{4}$ & $\frac{\sqrt{7}}{4}$ & $-\frac{3}{4}$ & $\frac{3}{4}$ & $0$ & $0$ & $0$ \\[1mm]
\hline
$\gamma$ & $0$ & $0$ & $1$ & $-1$ & $1$ & $-1$ & $0$ & $0$ & $2$ & $-2$ & $0$ & $0$ & $0$ \\[1mm]
\hline
$\delta$ & $\frac{8}{7}$ & $0$ & $2$ & $2$ & $2$ & $2$ & $4$
& $-\frac{4}{\sqrt{7}}$ & $0$ & $0$ & $\frac{4}{7}$ & $-\frac{4}{7}$
&
$8$ \\[1mm] \hline
\end{tabular}
}
\end{center}
{\it \caption{The scaling weights of the nine-dimensional fields. As
already anticipated, only three rescalings are independent since
they are subject to the following constraint:
$8\alpha-48\beta-18\gamma-9\delta=0$. As the scaling weight of the
Lagrangian $\mathcal{L}$ shows, $\beta$ and $\gamma$ belong to the
off-shell symmetries, whereas $\alpha$ and $\delta$ can be combined
into a trombone symmetry and an off-shell
symmetry.}\label{rescalings}}
\end{table}
The full field content consists of the following objects which
arrange themselves into irrep's of
$\mathbb{R}^{+}\,\times\,\textrm{SL}(2)$:
\be \textrm{9D :}\qquad \underbrace{e_{\mu}^{\phantom{\mu}a}\,,\,A_{\mu}\,,\,A_{\mu}{}^{i}\,,\,B_{\mu\nu}{}^{i}\,,\,C_{\mu\nu\rho}\,,\,\varphi\,,\,\tau=\,\chi\,+\,i\,e^{-\phi}}_{\textrm{bosonic dof's}}\,\,\,;\,\underbrace{\psi_\mu\,,\,\lambda\,,\,\tilde{\lambda}}_{\textrm{fermionic dof's}}\ ,\label{fields_9D} \ee%
where $\mu,\nu,\cdots$ denote nine-dimensional curved spacetime,
$a,b,\cdots$ nine-dimensional flat spacetime and $i,j,\cdots$
fundamental SL($2$) indices respectively.

The general deformations of this theory have been studied in detail
in ref.~\cite{FernandezMelgarejo:2011wx}, where both embedding
tensor deformations and gaugings of the trombone symmetry have been
considered. For the present scope we shall restrict ourselves to the
first ones. The latter ones would correspond to the additional mass
parameters $m_{\textrm{IIB}}$ and $(m_{11},m_{\textrm{IIA}})$ in
refs~\cite{Bergshoeff:2002nv, FernandezMelgarejo:2011wx}, which give
rise to theories without an action principle.

The vectors of the theory $\{A_{\mu}\,,\,{A_{\mu}}^{i}\}$ transform
in the $V^\prime\,=\,\textbf{1}_{(+4)}\,\oplus\,\textbf{2}_{(-3)}\,$
of $\mathbb{R}^{+}\,\times\,\textrm{SL}(2)\,$, where the
$\mathbb{R}^{+}$ scaling weights are included as well\footnote{The
$\mathbb{R}^{+}$ factor in the global symmetry is precisely the
combination
$\left(\frac{4}{3}\,\alpha\,-\,\frac{3}{2}\,\delta\right)\,$ of the
different rescalings introduced in ref.~\cite{Bergshoeff:2002nv}.}.
The resulting embedding tensor deformations live in the following
tensor product
\be \mathfrak{g}_{0}\otimes V=\textbf{1}_{(-4)}\,\oplus\, 2\,\cdot\,\textbf{2}_{(+3)}\,\oplus\,\textbf{3}_{(-4)}\,\oplus\,\textbf{4}_{(+3)}\ . \ee
The Linear Constraint (LC) projects out the $\textbf{4}_{(+3)}$, the
$\textbf{1}_{(-4)}$ and one copy of the $\textbf{2}_{(+3)}$ since
they would give rise to inconsistent deformations. As a consequence,
the consistent gaugings are parameterised by embedding tensor
components in the $\textbf{2}_{(+3)}\,\oplus\,\textbf{3}_{(-4)}$. We
will denote these allowed deformations by $\theta^{i}$ and
$\kappa^{(ij)}$.

The closure of the gauge algebra and the antisymmetry of the brackets impose the following Quadratic Constraints (QC)
\bea
\epsilon _{ij}\,\theta^{i}\,\kappa^{jk} &=&0\ ,\qquad\qquad
\textbf{2}_{(-1)}\label{quadratic constraints in 9D1}\\
\theta^{(i}\,\kappa^{jk)} &=&0\ .\qquad\qquad\,
\textbf{4}_{(-1)}\label{quadratic constraints in 9D2}
\eea

\subsubsection*{The $\mathbb{R}^{+}\,\times\,$SL($2$) orbits of
solutions to the QC}

The QC \eqref{quadratic constraints in 9D1} and \eqref{quadratic
constraints in 9D2} turns out to be very simple to solve; after
finding all the solutions, we studied the duality orbits,
\emph{i.e.} classes of those solutions which are connected via a
duality transformation. The resulting orbits of consistent gaugings
in this case are presented in table~\ref{orbits_max9}.

\begin{table}[h!]
\begin{center}
\scalebox{1}[1]{
\begin{tabular}{| c | c | c | c |}
\hline
\textrm{ID} & $\theta^{i}$ & ${\kappa}^{ij}$ &  gauging \\[1mm]
\hline \hline
$1$ & \multirow{3}{*}{$(0,0)$} &  diag($1,1$) &  SO($2$) \\[1mm]
\cline{1-1}\cline{3-4} $2$ & & diag($1,-1$) & SO($1,1$) \\[1mm]
\cline{1-1}\cline{3-4} $3$ & & diag($1,0$) & $\mathbb{R}^{+}_{\gamma}$ \\[1mm]
\hline \hline $4$ & $(1,0)$ & diag($0,0$) & $\mathbb{R}^{+}_{\beta}$ \\[1mm]
\hline
\end{tabular}
}
\end{center}
{\it \caption{All the U-duality orbits of consistent gaugings in maximal
supergravity in $D=9$. For each of them, the simplest representative
is given. The subscripts $\beta$ and $\gamma$ refer to the
rescalings summarised in table~\ref{rescalings}.}}
\label{orbits_max9}
\end{table}

\subsubsection*{Higher-dimensional geometric origin}

The four different orbits of maximal $D=9$ theory have the following higher-dimensional origin in terms of geometric compactifications  \cite{Bergshoeff:2002mb}:

\begin{itemize}

\item \textbf{Orbits 1 -- 3:}  These come from reductions of type IIB supergravity on a circle with an SL($2$) twist.

\item \textbf{Orbit 4:}  This can be obtained from a reduction of type IIA supergravity on a circle with the inclusion of an $\mathbb{R}^{+}_{\beta}$ twist.

\end{itemize}

\subsection{Orbits and origin of the $D=8$ maximal case}
\label{subsec:Max8}

\subsubsection*{Maximal $D=8$ gauged supergravity}

The maximal (ungauged) supergravity in $D=8$ \cite{Salam:1984ft} can
be obtained by reducing eleven-dimensional supergravity on a $T^3$.
The global symmetry group of this theory is
\be G_{0}\,=\,\textrm{SL}(2)\,\times\,\textrm{SL}(3)\ . \notag\ee
The full field content consists of the following objects which
arrange themselves into irrep's of
$\textrm{SL}(2)\,\times\,\textrm{SL}(3)$:
\be \textrm{8D :}\qquad \underbrace{e_{\mu}^{\phantom{\mu}a}\,,\,A_{\mu}{}^{\alpha m}\,,\,B_{\mu\nu m}\,,\,C_{\mu\nu\rho}\,,\,L_{m}^{\phantom{m}I}\,,\,\phi\,,\,\chi}_{\textrm{bosonic dof's}}\,\,\,;\,\underbrace{\psi_\mu\,,\,\chi_I}_{\textrm{fermionic dof's}}\ ,\label{fields_8D} \ee%
where $\mu,\nu,\cdots$ denote eight-dimensional curved spacetime,
$a,b,\cdots$ eight-dimensional flat spacetime, $m,n,\cdots$
fundamental SL($3$), $I,J,\cdots$ fundamental SO($3$) and
$\alpha,\beta,\cdots$ fundamental SL($2$) indices respectively. The
six vector fields $A_{\mu}{}^{\alpha m}$ in \eqref{fields_8D}
transform in the $V'=\left( \textbf{2},\textbf{3}^\prime\right)$.
There are eleven group generators, which can be expressed in the
adjoint representation $\mathfrak{g}_{0}$.

The embedding tensor $\Theta $ then lives in the representation
$\mathfrak{g} _{0}\,\otimes\,V$, which can be decomposed into
irreducible representations as
\be \mathfrak{g}_{0}\otimes V=2\,\cdot\left(
\textbf{2},\textbf{3}\right) \oplus \left(
\textbf{2},\textbf{6}^\prime\right) \oplus  \left(
\textbf{2},\textbf{15}\right) \oplus \left(
\textbf{4},\textbf{3}\right)\,. \ee
The LC restricts the embedding tensor to the
$\left( \textbf{2},\textbf{3}\right) \oplus \left(
\textbf{2},\textbf{6}^\prime\right) $ \cite{Weidner:2006rp}. It is
worth noticing that there are two copies of the $\left(
\textbf{2},\textbf{3} \right) $ irrep in the above composition; the
LC imposes a relation between them
\cite{Samtleben:2008pe}. This shows that, for consistency, gauging
some SL($2$) generators implies the necessity of gauging some
SL($3$) generators as well. Let us denote the allowed embedding
tensor irrep's by $\xi _{\alpha m}$ and $f_{\alpha}{}^{(mn)}$
respectively.

The quadratic constraints (QC) then read \cite{Dani:2008,
deRoo:2011fa}
\bea
\epsilon ^{\alpha\beta}\,\xi _{\alpha p}\xi _{\beta q} &=&0
\text{ ,}\qquad\qquad
\left( \textbf{1},\textbf{3}^\prime\right)\label{quadratic constraints in 8D1}\\%
f_{(\alpha}{}^{np}\xi _{\beta)p} &=&0 \text{ ,} \qquad\qquad \left(
\textbf{3},\textbf{3}^\prime\right)\label{quadratic constraints in 8D2}\\%
\epsilon ^{\alpha\beta}\left(\epsilon
_{mqr}f_{\alpha}{}^{qn}f_{\beta}{}^{rp}+f_{\alpha}{}^{np}\xi _{\beta
m}\right)
&=&0\text{ .} \qquad\left( \textbf{1},\textbf{3}^\prime\right)\oplus\left( \textbf{1},\textbf{15}\right)\label{quadratic constraints in 8D3} %
\eea
Any solution to the QC \eqref{quadratic constraints in 8D1},
\eqref{quadratic constraints in 8D2} and \eqref{quadratic
constraints in 8D3} specifies a consistent gauging of a subgroup of
SL($2)\,\times\,$SL($3$) where the corresponding generators are
given by
\bea \label{gauge_gen_D=8} {\left(X_{\alpha
m}\right)_{\beta}}^{\gamma} &=& \delta_{\alpha}^{\gamma}\,\xi_{\beta
m}\,-\,\frac{1}{2}\,\delta_{\beta}^{\gamma}\,\xi_{\alpha m}\ ,\\
{\left(X_{\alpha m}\right)_{n}}^{p} &=&
\epsilon_{mnq}\,{f_{\alpha}}^{qp}\,-\,\frac{3}{4}\,\left(\delta_{m}^{p}\,\xi_{\alpha
n}\,-\,\frac{1}{3}\,\delta_{n}^{p}\,\xi_{\alpha m}\right)\ . \eea

\subsubsection*{The SL($2$)$\,\times\,$SL($3$) orbits of solutions to
the QC}

We exploited an algebraic geometry tool called the
Gianni-Trager-Zacharias (GTZ) algorithm \cite{GTZ}. This algorithm
has been computationally implemented by the \textsc{\,Singular\,}
project \cite{DGPS} and it consists in the primary decomposition of
ideals of polynomials. After finding all the solutions to the QC by
means of the algorithm mentioned above, one has to group together
all the solutions which are connected through a duality
transformation, thus obtaining a classification of such solutions in
terms of duality orbits. The resulting orbits of consistent
gaugings\footnote{Recently, also the possible vacua of the different
theories have been analysed \cite{deRoo:2011fa}. It was found that
only  {\bf orbit 3} has maximally symmetric vacua.} in this case are
presented in table~\ref{orbits_max8}.

\begin{table}[h!]
\begin{center}
\scalebox{1}[1]{
\begin{tabular}{| c | c | c | c | c | c |}
\hline
\textrm{ID} & ${f_{+}}^{mn}$ & ${f_{-}}^{mn}$ & $\xi_{+m}$ & $\xi_{-m}$ &  gauging \\[1mm]
\hline \hline
$1$ & diag($1,1,1$) & \multirow{5}{*}{$\textrm{diag}(0,0,0)$} & \multirow{5}{*}{$(0,0,0)$} & \multirow{5}{*}{$(0,0,0)$} &  SO($3$) \\[1mm]
\cline{1-2}\cline{6-6} $2$ & diag($1,1,-1$) & & & &  SO($2,1$) \\[1mm]
\cline{1-2}\cline{6-6} $3$ & diag($1,1,0$) & & & &  ISO($2$) \\[1mm]
\cline{1-2}\cline{6-6} $4$ & diag($1,-1,0$) & & & &  ISO($1,1$) \\[1mm]
\cline{1-2}\cline{6-6} $5$ & diag($1,0,0$) & & & &  CSO($1,0,2$) \\[1mm]
\hline \hline $6$ & diag($0,0,0$) & diag($0,0,0$) & $(1,0,0)$ &
$(0,0,0)$ & Solv$_{2}\,\times\,$Solv$_{3}$\\[1mm]
\hline \hline $7$ & diag($1,1,0$) &
\multirow{3}{*}{$\textrm{diag}(0,0,0)$} & \multirow{3}{*}{$(0,0,1)$}
& \multirow{3}{*}{$(0,0,0)$} & \multirow{3}{*}{Solv$_{2}\,\times\,$Solv$_{3}$}\\[1mm]
\cline{1-2} $8$ & diag($1,-1,0$) & & & &   \\[1mm]
\cline{1-2} $9$ & diag($1,0,0$) & & & &   \\[1mm]
\hline \hline $10$ & diag($1,-1,0$) & \scalebox{0.7}[0.7]{$\left(\begin{array}{ccc}1 & 1 & 0\\
1 & 1 & 0\\ 0 & 0 & 0\end{array}\right)$} & $\frac{2}{9}(0,0,1)$ &
$(0,0,0)$ & Solv$_{2}\,\times\,$SO$(2)\,\ltimes\,$Nil$_{3}(2)$\\[1mm]

\hline
\end{tabular}
}
\end{center}
{\it \caption{All the U-duality orbits of consistent gaugings in maximal
supergravity in $D=8$. For each of them, the simplest representative
is given. We denote by  Solv$_{2}\,\subset\,$SL($2$) and
Solv$_{3}\,\subset\,$SL($3$) a solvable algebra of dimension 2 and 3
respectively. To be more precise, Solv$_{2}$ identifies the Borel
subgroup of SL($2$) consisting of $2\times 2$ upper-triangular
matrices. Solv$_{3}$, instead, is a Bianchi type V algebra.}}
\label{orbits_max8}
\end{table}

\subsubsection*{Higher-dimensional geometric origin}

\begin{itemize}
\item \textbf{Orbits 1 -- 5:}  These stem from reductions of eleven-dimensional supergravity on a three-dimensional group manifold of type  A in the Bianchi classification \cite{Bergshoeff:2003ri}. The special case in orbit 1 corresponds to a reduction over an SO($3$) group manifold and it was already studied in ref.~\cite{Salam:1984ft}.

\item \textbf{Orbit 6:}  This can be obtained from a reduction of maximal nine-dimensional supergravity on a circle with the inclusion of an  $\mathbb{R}^{+}$ twist inside the global symmetry group.

\item \textbf{Orbits 7 -- 9:}  These can come from the same reduction from $D=9$ but upon inclusion of a more general $\mathbb{R}^{+}\,\times\,\textrm{SL}(2)$ twist.

\item \textbf{Orbit 10:}  This orbit seems at first sight more complicated to be obtained from a dimensional reduction owing to its non-trivial SL($2$) angles. Nevertheless, it turns out that one can land on this orbit by compactifying type IIB supergravity on a circle with an SL($2$) twist and then further reducing on another circle with $\mathbb{R}^{+}\,\times\,\textrm{SL}(2)$ twist given by the residual little group leaving invariant the intermediate nine-dimensional deformation.
\end{itemize}

\subsection*{Remarks on the $D=7$ maximal case}
\label{subsec:Max7}

The general deformations of the maximal theory in $D=7$ are
constructed and presented in full detail in
ref.~\cite{Samtleben:2005bp}. For the present aim we only summarise
here a few relevant facts.

The global symmetry group of the theory is SL($5$). The vector
fields $A_{\mu}{}^{MN}=A_{\mu}{}^{[MN]}$ transform in the
\textbf{10}$^\prime$ of SL($5$), where we denote by $M$ a
fundamental SL($5$) index. The embedding tensor $\Theta$ takes
values in the following irreducible components
\be
\textbf{10}\otimes\textbf{24}\,=\,\textbf{10}\oplus\textbf{15}\oplus\textbf{40}^\prime\oplus
\textbf{175}\,.\ee
The LC restricts the embedding tensor to the $
\textbf{15}\,\oplus\,\textbf{40}^\prime$, which can be parameterised
by the following objects
\be Y_{(MN)}\,,\qquad\textrm{and}\qquad
Z^{[MN],P}\quad\textrm{with}\quad Z^{[MN,P]}=0\ . \ee
The generators of the gauge algebra can be written as follows
\be
{\left(X_{MN}\right)_P}^Q\,=\,\delta_{[M}^Q\,Y_{N]P}\,-\,2\,\epsilon_{MNPRS}\,Z^{RS,Q}\,,
\label{gen_max}\ee
or, identically, if one wants to express them in the $\textbf{10}$,
\be
{\left(X_{MN}\right)_{PQ}}^{RS}\,=\,2\,{\left(X_{MN}\right)_{[P}}^{[R}\,\,\delta_{Q]}^{S]}\,.
\label{gen_max10}\ee
The closure of the gauge algebra and the antisymmetry of the
brackets imply the following QC
\be
Y_{MQ}\,Z^{QN,P}\,+\,2\,\epsilon_{MRSTU}\,Z^{RS,N}\,Z^{TU,P}\,=\,0\
, \label{QC_max7}\ee
which have different irreducible pieces in the
$\textbf{5}^\prime\,\oplus\,\textbf{45}^\prime\,\oplus\,\textbf{70}^\prime$.
Unfortunately, in this case, both the embedding tensor deformations
and the quadratic constraints reach a level of complexity that makes an exhaustive and general analysis difficult.
 Such analysis lies beyond the scope of our work.

\section{T-duality orbits of half-maximal supergravities}
\label{sec:T_Dualitites}

After the previous section on maximal supergravities, we turn our
attention to theories with half-maximal supersymmetry. In
particular, in this section we will classify the orbits under
T-duality of all gaugings of half-maximal supergravity. We will only
consider the theories with duality groups $\mathbb{R}^+ \times
\textrm{SO}(d,d)$ in $D= 10-d$, which places a restriction on the
number of vector multiplets. For these theories we will classify all
duality orbits, and find a number of non-geometric orbits.
Furthermore, we demonstrate that double field theory does yield a
higher-dimensional origin for all of them.

Starting from $D=10$ half-maximal supergravity without vector
multiplets, it can be seen that there is no freedom to deform this
theory, rendering this case trivial. In $D=9$, instead, we have the
possibility of performing an Abelian gauging inside
$\mathbb{R}^{+}\,\times\,$SO($1,1$), which will depend on one
deformation parameter. However, this is precisely the parameter that
one expects to generate by means of a twisted reduction from $D=10$.
This immediately tells us that non-geometric fluxes do not yet
appear in this theory. In order to find the first non-trivial case,
we will have to consider the $D=8$ case.

\subsection{Orbits and origin of the $D=8$ half-maximal case}
\label{subsec:Half_Max8}

\subsubsection*{Half-maximal $D=8$ gauged supergravity}

Half-maximal supergravity in $D=8$ is related to the maximal theory
analysed in the previous section by means of a $\mathbb{Z}_{2}$
truncation. The action of such a $\mathbb{Z}_{2}$ breaks
$\textrm{SL}(2)\times\textrm{SL}(3)$ into
$\mathbb{R}^{+}\times\textrm{SL}(2)\times\textrm{SL}(2)$, where
$\textrm{SL}(2)\times\textrm{SL}(2)=\textrm{O}(2,2)$ can be
interpreted as the T-duality group in $D=8$ as shown in
table~\ref{dualities}. The embedding of
$\mathbb{R}^{+}\times\textrm{SL}(2)$ inside SL($3$) is unique and it
determines the following branching of the fundamental representation
\bea
\textbf{3}\,\,&\longrightarrow&\,\,\textbf{1}_{(+2)}\,\oplus\,\textbf{2}_{(-1)}\ ,\notag\\
m\,&\longrightarrow&\,\,(\bullet\,,\,i)\ ,\notag \eea
where the $\mathbb{R}^{+}$ direction labeled by $\bullet$ is parity
even, whereas $i$ is parity odd, such as the other SL($2$) index
$\alpha$. In the following we will omit all the $\mathbb{R}^{+}$ weights since they do not play any role in the truncation.

The embedding tensor of the maximal theory splits in the following
way
\bea
(\textbf{2},\textbf{3})\,\,&\longrightarrow&\,\,\xcancel{(\textbf{2},\textbf{1})}\,\oplus\,(\textbf{2},\textbf{2})\ ,\notag\\
(\textbf{2},\textbf{6}^{\prime})\,\,&\longrightarrow&\,\,\xcancel{(\textbf{2},\textbf{1})}\,\oplus\,(\textbf{2},\textbf{2})\,\oplus\,\xcancel{(\textbf{2},\textbf{3})}\
,\notag \eea
where all the crossed irrep's are projected out because of
$\mathbb{Z}_2$ parity. This implies that the consistent embedding
tensor deformations of the half-maximal theory can be described by
two objects which are doublets with respect to both SL($2$)'s. Let
us denote them by $a_{\alpha i}$ and $b_{\alpha i}$. This statement
is in perfect agreement with the Kac-Moody analysis performed in
ref.~\cite{Bergshoeff:2007vb}. The explicit way of embedding
$a_{\alpha i}$ and $b_{\alpha i}$ inside $\xi _{\alpha m}$ and
$f_{\alpha}{}^{mn}$ is given by
\bea
{f_{\alpha}}^{i\bullet}&=&{f_{\alpha}}^{\bullet i}\,=\,\epsilon^{ij}\,a_{\alpha j}\ , \label{ET_Half_Max81}\\[2mm]
\xi _{\alpha i}&=&4\,b _{\alpha i}\ .\label{ET_Half_Max82} \eea

The QC given in \eqref{quadratic constraints in 8D1},
\eqref{quadratic constraints in 8D2} and \eqref{quadratic
constraints in 8D3} are decomposed according to the following
branching
\bea
(\textbf{1},\textbf{3}^{\prime})\,\,&\longrightarrow&\,\,(\textbf{1},\textbf{1})\,\oplus\,\xcancel{(\textbf{1},\textbf{2})}\ ,\notag\\
(\textbf{3},\textbf{3}^{\prime})\,\,&\longrightarrow&\,\,(\textbf{3},\textbf{1})\,\oplus\,\xcancel{(\textbf{3},\textbf{2})}\ ,\notag\\
(\textbf{1},\textbf{15})\,\,&\longrightarrow&\,\,(\textbf{1},\textbf{1})\,\oplus\,\xcancel{2\,\cdot\,(\textbf{1},\textbf{2})}\,\oplus\,2\,\cdot\,(\textbf{1},\textbf{3})\,\oplus\,\xcancel{(\textbf{1},\textbf{4})}\
.\notag\eea
As a consequence, one expects the set of $\mathbb{Z}_{2}$ even QC to
consist of 3 singlets, a $(\textbf{3},\textbf{1})$ and 2 copies of
the $(\textbf{1},\textbf{3})$. By plugging \eqref{ET_Half_Max81} and
\eqref{ET_Half_Max82} into \eqref{quadratic constraints in 8D1},
\eqref{quadratic constraints in 8D2} and \eqref{quadratic
constraints in 8D3}, one finds
\bea \epsilon ^{\alpha\beta}\,\epsilon^{ij}\,b_{\alpha i}\,b_{\beta
j}&=&0\ ,\qquad\qquad
\left(\textbf{1},\textbf{1}\right)\label{QC_Half_Tot_81}\\
\epsilon ^{\alpha\beta}\,\epsilon^{ij}\,a_{\alpha i}\,b_{\beta
j}&=&0\ ,\qquad\qquad
\left(\textbf{1},\textbf{1}\right)\label{QC_Half_Tot_82}\\
\epsilon ^{\alpha\beta}\,\epsilon^{ij}\,a_{\alpha i}\,a_{\beta
j}&=&0\ ,\qquad\qquad
\left(\textbf{1},\textbf{1}\right)\label{QC_Half_Tot_83}\\
\epsilon^{ij}\,a_{(\alpha i}\,b_{\beta) j}&=&0\ ,\qquad\qquad
\left(\textbf{3},\textbf{1}\right)\label{QC_Half_Tot_84}\\
\epsilon ^{\alpha\beta}\,a_{\alpha (i}\,b_{\beta j)}&=&0\
.\qquad\qquad
\left(\textbf{1},\textbf{3}\right)\label{QC_Half_Tot_85} \eea
With respect to what we expected from group theory, we seem to be
finding a $(\textbf{1},\textbf{3})$ less amongst the even QC. This
could be due to the fact that $\mathbb{Z}_{2}$ even QC can be
sourced by quadratic expressions in the odd embedding tensor
components that we truncated away. After the procedure of turning
off all of them, the two $(\textbf{1},\textbf{3})$'s probably
collapse to the same constraint or one of them vanishes directly.

The above set of QC characterises the consistent gaugings of the
half-maximal theory which are liftable to the maximal theory, and
hence they are more restrictive than the pure consistency
requirements of the half-maximal theory. In order to single out only
these we need to write down the expression of the gauge generators
and impose the closure of the algebra. The gauge generators in the
$(\textbf{2},\textbf{2})$ read
\be \label{Gen_Half_max8} {\left(X_{\alpha i}\right)_{\beta
j}}^{\gamma k} =
\frac{1}{2}\,\delta^{\gamma}_{\beta}\,\epsilon_{ij}\,\epsilon^{kl}\,a_{\alpha
l} \,+\, \delta^{\gamma}_{\alpha}\,\delta^{k}_{j}\,b_{\beta i} \,-\,
\frac{3}{2}\,\delta^{\gamma}_{\beta}\,\delta^{k}_{i}\,b_{\alpha j}
\,+\,
\frac{1}{2}\,\delta^{\gamma}_{\beta}\,\delta^{k}_{j}\,b_{\alpha i}
\,+\, \epsilon_{\alpha \beta}\,\epsilon^{\gamma
\delta}\,\delta^{k}_{j}\,b_{\delta i}\ . \ee
The closure of the algebra generated by \eqref{Gen_Half_max8}
implies the following QC
\bea \epsilon ^{\alpha\beta}\,\epsilon^{ij}\,\left(a_{\alpha
i}\,a_{\beta j}\,-\,b_{\alpha i}\,b_{\beta j}\right)&=&0\
,\qquad\qquad
\left(\textbf{1},\textbf{1}\right)\label{QC_Half_81}\\
\epsilon ^{\alpha\beta}\,\epsilon^{ij}\,\left(a_{\alpha i}\,b_{\beta
j}\,+\,b_{\alpha i}\,b_{\beta j}\right)&=&0\ ,\qquad\qquad
\left(\textbf{1},\textbf{1}\right)\label{QC_Half_82}\\
\epsilon^{ij}\,a_{(\alpha i}\,b_{\beta) j}&=&0\ ,\qquad\qquad
\left(\textbf{3},\textbf{1}\right)\label{QC_Half_83}\\
\epsilon ^{\alpha\beta}\,a_{\alpha (i}\,b_{\beta j)}&=&0\
.\qquad\qquad \left(\textbf{1},\textbf{3}\right)\label{QC_Half_84}
\eea

To facilitate the mapping of gaugings $a_{\alpha i}$ and $b_{\alpha i}$ with the more familiar $f_{ABC}$ and $\xi_A$ in the DFT language, we have written a special section in the appendix \ref{appendix_B}. The mapping is explicitly given in \eqref{X2f_D=8}.

\subsubsection*{The O($2,2$) orbits of solutions to the QC}

After solving the QC given in \eqref{QC_Half_81},
\eqref{QC_Half_82}, \eqref{QC_Half_83} and \eqref{QC_Half_84} again
with the aid of \textsc{\,Singular\,}, we find a 1-parameter
family of T-duality orbits plus two discrete ones. The results are
all collected in table~\ref{orbits_half_max8}.

\begin{table}[h!]
\begin{center}
\scalebox{1}[1]{
\begin{tabular}{| c | c | c | c |}
\hline
\textrm{ID} & $a_{\alpha i}$ & $b_{\alpha i}$ & gauging \\[1mm]
\hline \hline
$1$ & diag($\,\cos\alpha,0$) & diag($\,\sin\alpha,0$) &  Solv$_{2}\,\times\,$SO($1,1$) \\[1mm]
\hline \hline
$2$ & diag($1,1$) & diag($-1,-1$) &  \multirow{2}{*}{SL$(2)\,\times\,$SO($1,1$)} \\[1mm]
\cline{1-3}$3$ & diag($1,-1$) & diag($-1,1$) &  \\[1mm]
\hline
\end{tabular}
}
\end{center}
{\it \caption{All the T-duality orbits of consistent gaugings in
half-maximal supergravity in $D=8$. For each of them, the simplest
representative is given. Solv$_{2}$ refers again to the solvable
subgroup of SL($2$) as already explained in the caption of
table~\ref{orbits_max8}.} \label{orbits_half_max8}}
\end{table}

\subsubsection*{Higher-dimensional geometric origin}

The possible higher-dimensional origin of the three different orbits is as follows:

\begin{itemize}

\item \textbf{Orbit 1:} This orbit can be obtained by performing a
two-step reduction of type I supergravity. In the first step, by
reducing a circle, we can generate an
$\mathbb{R}^{+}\,\times\,$SO($1,1$) gauging of half-maximal $D=9$
supergravity. Subsequently, we reduce such a theory again on a
circle with the inclusion of a new twist commuting with the previous
deformation. Also, these orbits include a non-trivial $\xi_A$ gauging, so we will not address it from a DFT perspective.

\item \textbf{Orbits 2 -- 3:} These do not seem to have any obvious geometric
higher-dimensional origin in supergravity. In fact, they do not
satisfy the extra constraints \eqref{Extra_f}, so one can only hope
to reproduce them from truly doubled twist orbits in DFT.

\end{itemize}
Therefore we find that, while the half-maximal orbits in $D = 9$ all have a known geometric higher-dimensional origin, this is not the case for the latter two orbits in $D = 8$. We have finally detected the first signals of
non-geometric orbits.

\subsubsection*{Higher-dimensional DFT origin}

As mentioned, the {\bf orbits 2} and {\bf 3} lack of a clear
higher-dimensional origin. Here we would like to provide a
particular twist matrix giving rise to these gaugings. We chose to
start in the cartesian framework, and propose the following form for
the SO$(2,2)$ twist matrix \be U = \begin{pmatrix} 1& 0 & 0 & 0 \\ 0
&
\cosh (m\,y^1 + n \, \tilde y_1) & 0 & \sinh (m\,y^1 + n \, \tilde y_1) \\
0& 0& 1& 0\\ 0 & \sinh (m\,y^1 + n \, \tilde y_1) & 0 & \cosh (m\,y^1
+ n \, \tilde y_1) \end{pmatrix}\ .  \vspace{2mm}\ee

\noindent This is in fact an element of $\textrm{SO}(1,1)$ lying in the
directions ($\tilde y_2, y^2$), fibred over the double torus
($\tilde y_1 , y^1 $). Here, the coordinates are written in the
cartesian formulation, so we must rotate this in order to make
contact with the light-cone case.

For this twist matrix, the weak and strong constraints in the
light-cone formulation read $(m+n) (m-n) = 0$, while the QC are
always satisfied. The gaugings are constant, and when written in
terms of $a_{\alpha i}$ and $b_{\alpha i}$ we find \be a_{\alpha i}
= - b_{\alpha i } ={\rm diag} \left(-\frac{m+n}{2\,\sqrt{2}} ,\
\frac{m-n}{2\,\sqrt{2}}\right)\ , \ee so {\bf orbit 2} is obtained
by choosing $m = 0$, $n = -2\,\sqrt{2}$, and {\bf orbit 3} by
choosing $m = - 2\,\sqrt{2}$, $n = 0$. Notice that in both cases the
twist orbit is truly doubled, so we find the first example of an
orbit of gaugings without a clear supergravity origin, that finds an
uplift to DFT in a truly doubled background.

\subsection{Orbits and origin of the $D=7$ half-maximal case}
\label{subsec:Half_Max7}

\subsubsection*{Half-maximal $D=7$ gauged supergravity}

A subset of half-maximal gauged supergravities is obtained from the
maximal theory introduced in section~\ref{subsec:Max7} by means of a
$\mathbb{Z}_2$ truncation. Thus, we will in this section perform
this truncation and carry out the orbit analysis in the half-maximal
theory. As we already argued before, this case is not only simpler,
but also much more insightful from the point of view of
understanding T-duality in gauged supergravities and its relation to
DFT.

The action of our $\mathbb{Z}_2$ breaks\footnote{The $\mathbb{Z}_2$
element with respect to which we are truncating is the following
USp($4)\,=\,$SO($5$) element \be
\alpha\,=\,\left(\begin{array}{cc}\mathds{1}_{2} & 0\\ 0 &
-\mathds{1}_{2}\end{array}\right) \notag\ee projecting out half of
the supercharges.} SL($5$) into $\mathbb{R}^+\,\times\,$SL($4$). Its
embedding inside SL($5$) is unique and it is such that the
fundamental representation splits as follows
\be
\textbf{5}\,\,\longrightarrow\,\,\textbf{1}_{(+4)}\,\oplus\,\textbf{4}_{(-1)}\,.
\ee
After introducing the following notation for the indices in the $\mathbb{R}^+$ and in the SL($4$) directions
\be M\,\,\longrightarrow\,\,(\,\diamond\,,\,m)\,, \ee
we assign an even parity to the $\diamond$ direction and odd parity
to $m$ directions.

The embedding tensor of the maximal theory splits according to
\bea
\textbf{15}&\longrightarrow & \textbf{1}\,\oplus\,\xcancel{\textbf{4}}\,\oplus\,\textbf{10}\,,\\[2mm]
\textbf{40}^\prime&\longrightarrow &\xcancel{\textbf{4}^\prime}\,\oplus\,\textbf{6}\,\oplus\,\textbf{10}^\prime\,\oplus\,\xcancel{\textbf{20}}\,,
\eea
where again, as in section~\ref{subsec:Half_Max8}, all the crossed
irrep's are projected out because of $\mathbb{Z}_2$ parity. This
implies that the embedding tensor of the half-maximal theory lives
in the
$\textbf{1}\,\oplus\,\textbf{6}\,\oplus\,\textbf{10}\,\oplus\,\textbf{10}^\prime$
and hence it is described by the following objects
\be
\theta\,\,,\,\,\xi_{[mn]}\,\,,\,\,M_{(mn)}\,\,,\,\,\tilde{M}^{(mn)}\,.
\label{Theta_half} \ee
This set of deformations agrees with the decomposition
$\textrm{D}_8^{+++}\,\rightarrow\,\textrm{A}_3\,\times\,\textrm{A}_6$
given in ref.~\cite{Bergshoeff:2007vb}. The objects in
\eqref{Theta_half} are embedded in $Y$ and $Z$ in the following way
\bea
Y_{\diamond\,\diamond}&=&\theta\,, \label{ExprY1}\\[2mm]
Y_{mn}&=&\frac{1}{2}\,M_{mn}\,,\label{ExprY2}\\[3mm]
Z^{mn,\,\diamond}&=&\frac{1}{8}\,\xi^{mn}\,,\label{ExprZ1}\\[2mm]
Z^{m\,\diamond,n}&=&-Z^{\diamond\,m,n}\,=\,\frac{1}{16}\,\tilde{M}^{mn}\,+\,\frac{1}{16}\,\xi^{mn}\,,\label{ExprZ2}
\eea
where for convenience we defined $\xi^{mn}\,=\, \frac{1}{2}\,\epsilon^{mnpq}\,\xi_{pq}$.

Now we will obtain the expression of the gauge generators of the
half-maximal theory by plugging the expressions \eqref{ExprY1} --
\eqref{ExprZ2} into \eqref{gen_max}. We find
\be
{\left(X_{mn}\right)_p}^q\,=\,\frac{1}{2}\,\delta_{[m}^q\,M_{n]p}\,-\,\frac{1}{4}\,\epsilon_{mnpr}\,\left(\tilde{M}\,+\,
\xi\right)^{rq}\, \,, \label{gen_half-max}\ee
which extends the expression given in  ref.~\cite{Roest:2009tt} by
adding an antisymmetric part to $\tilde{M}$ proportional to $\xi$.
Note that the $\xi$ term is also the only one responsible for the
trace of the gauge generators which has to be non-vanishing in order
to account for $\mathbb{R}^+$ gaugings.

The presence of such a term in the expression \eqref{gen_half-max}
has another consequence: the associated structure constants that one
writes by expressing the generators in the $\textbf{6}$
${\left(X_{mn}\right)_{pq}}^{rs}$ will not be automatically
antisymmetric in the exchange between $mn$ and $pq$. This implies
the necessity of imposing the antisymmetry by means of some extra
QC\footnote{The QC which
ensure the antisymmetry of the gauge brackets are given by \\
${\left(X_{mn}\right)_{pq}}^{rs}\,X_{rs}\,+\,(mn\,\leftrightarrow\,pq)\,=\,0$,
where $X$ is given in an arbitrary representation.}.

The QC of the maximal theory are branched into
\bea
\textbf{5}^\prime&\longrightarrow & \textbf{1}\,\oplus\,\xcancel{\textbf{4}^\prime}\,,\\[2mm]
\textbf{45}^\prime&\longrightarrow &\xcancel{\textbf{4}}\,\oplus\,\textbf{6}\,\oplus\,\textbf{15}\,\oplus\,\xcancel{\textbf{20}}\,,\\[2mm]
\textbf{70}^\prime&\longrightarrow &\textbf{1}\,\oplus\,\xcancel{\textbf{4}}\,\oplus\,\xcancel{\textbf{4}^\prime}\,\oplus\,\textbf{10}^\prime\,\oplus\,\textbf{15}\,\oplus\,\xcancel{\textbf{36}^\prime}\,.
\eea
By substituting the expressions \eqref{ExprY1} -- \eqref{ExprZ2}
into the QC \eqref{QC_max7}, one finds
\bea
\theta\,\xi_{mn}&=&0\,,\qquad \,\,\,\,\,\,(\textbf{6}) \label{theta_xi}\\[2mm]
\left(\tilde{M}^{mp}\,+\,  \xi^{mp}\right)\, M_{pq}&=&0\,,\qquad ( \textbf{1}\,\oplus\,\textbf{15})\label{Q_Qtilde}\\[2mm]
M_{mp}\,\xi^{pn}\,-\,\xi_{mp}\,\left(\tilde{M}^{pn}\,+\,\xi^{pn}\right)&=&0\,,\qquad ( \textbf{1}\,\oplus\,\textbf{15})\label{Q_xi}\\[2mm]
\theta\,\tilde{M}^{mn}&=&0\,.\qquad \,\,\,\,\,(\textbf{10}^\prime)
\label{theta_Qtilde} \eea
Based on the Kac-Moody analysis performed in
ref.~\cite{Bergshoeff:2007vb}, the QC constraints of the
half-maximal theory should only impose conditions living in the
$\textbf{1}\,\oplus\,\textbf{6}\,\oplus\,\textbf{15}\,\oplus\,\textbf{15}$.
The problem is then determining which constraint in the $\textbf{1}$
is already required by the half-maximal theory and which is not.

By looking more carefully at the constraints \eqref{theta_xi} --
\eqref{theta_Qtilde}, we realise that the traceless part of
\eqref{Q_Qtilde} exactly corresponds to the Jacobi identities that
one gets from the closure of the algebra spanned by the generators
\eqref{gen_half-max}, whereas the full \eqref{Q_xi} has to be
imposed to ensure antisymmetry of the gauge brackets. Since there is
only one constraint in the  $\textbf{6}$, we do not have ambiguities
there\footnote{We would like to stress that the parameter $\theta$
within the half-maximal theory is a consistent deformation, but it
does not correspond to any gauging and hence QC involving it cannot
be derived as Jacobi identities or other consistency constraints
coming from the gauge algebra.}.

We are now able to write down the set of QC of the half-maximal theory:
\bea
\theta\,\xi_{mn}&=&0\,,\qquad \,(\textbf{6}) \label{QC1}\\[2mm]
\left(\tilde{M}^{mp}\,+\,  \xi^{mp}\right)\, M_{pq}\,-\,\frac{1}{4}\,\left(\tilde{M}^{np}\,M_{np}\right)\,\delta_q^m&=&0\,,\qquad (\textbf{15})\label{QC2}\\[2mm]
M_{mp}\,\xi^{pn}\,+\,\xi_{mp}\,\tilde{M}^{pn}&=&0\,,\qquad (\textbf{15})\label{QC3}\\[2mm]
\epsilon^{mnpq}\,\xi_{mn}\,\xi_{pq}&=&0\,.\qquad \,\,\,(
\textbf{1})\label{QC4} \eea
We are not really able to confirm whether (\ref{QC1}) is part of the
QC of the half-maximal theory, in the sense that there appears a
top-form in the \textbf{6} from the $\textrm{D}_{8}^{+++}$
decomposition but it could either be a tadpole or a QC. This will
however not affect our further discussion, in that we only consider
orbits of gaugings in which $\theta=0$. The extra QC required in
order for the gauging to admit an uplift to maximal supergravity are
\bea
\tilde{M}^{mn}\,M_{mn}&=&0\,,\qquad \,(\textbf{1}) \label{extra1}\\[2mm]
\theta\,\tilde{M}^{mn}&=&0\,.\qquad (\textbf{10}^\prime)
\label{extra2} \eea

\subsubsection*{The O($3,3$) orbits of solutions to the QC in the
$\textbf{10}\,\oplus\,\textbf{10}^\prime$}

The aim of this section is to solve the constraints summarised in
\eqref{QC1}, \eqref{QC2}, \eqref{QC3} and \eqref{QC4}. We will start
by considering the case of gaugings only involving the
$\textbf{10}\,\oplus\,\textbf{10}^\prime$. This restriction is
motivated by flux compactification, as we will try to argue later
on.

The only non-trivial QC are the following
\be
\tilde{M}^{mp}\,M_{pn}-\frac{1}{4}\left(\tilde{M}^{pq}\,M_{pq}\right)\,\delta^m_n\,=\,0\,,
\label{QCQQtilde}\ee
which basically implies that the matrix product between $M$ and
$\tilde{M}$, which in principle lives in the
$\textbf{1}\,\oplus\,\textbf{15}$, has to be pure trace. We made use
of a GL($4$) transformation in order to reduce $M$ to pure
signature; as a consequence, the QC \eqref{QCQQtilde} imply that
$\tilde{M}$ is diagonal as well \cite{Dibitetto:2010rg}. This
results in a set of eleven 1-parameter orbits\footnote{We would like
to point out that the extra discrete generator $\eta$ of O($3,3$)
makes sure that, given a certain gauging with $M$ and $\tilde{M}$,
it lies in the same orbit as its partner with the role of $M$ and
$-\tilde{M}$ interchanged.} of solutions to the QC which are given
in table~\ref{orbits_halfmax7}.

\begin{table}[h!]
\begin{center}
\scalebox{1}[1]{
\begin{tabular}{| c | c | c | c | c |}
\hline
\textrm{ID} & $M_{mn}/\,\cos\alpha\,$ & $\tilde{M}^{mn}/\,\sin\alpha\,$ & range of $\alpha$ & gauging \\[1mm]
\hline \hline
$1$ & diag($1,1,1,1$) & diag($1,1,1,1$) & $-\frac{\pi}{4}\,<\,\alpha\,\le\,\frac{\pi}{4}$ & $\left\{\begin{array}{cc}\textrm{SO}($4$)\ , & \alpha\,\ne\,\frac{\pi}{4}\ ,\\ \textrm{SO}(3)\ , & \alpha\,=\,\frac{\pi}{4}\ .\end{array}\right.$\\[4mm]
\hline
$2$ & diag($1,1,1,-1$) & diag($1,1,1,-1$) & $-\frac{\pi}{4}\,<\,\alpha\,\le\,\frac{\pi}{4}$ & SO($3,1$)\\[1mm]
\hline
$3$ & diag($1,1,-1,-1$) & diag($1,1,-1,-1$) & $-\frac{\pi}{4}\,<\,\alpha\,\le\,\frac{\pi}{4}$ & $\left\{\begin{array}{cc}\textrm{SO}($2,2$)\ , & \alpha\,\ne\,\frac{\pi}{4}\ ,\\ \textrm{SO}(2,1)\ , & \alpha\,=\,\frac{\pi}{4}\ .\end{array}\right.$\\[2mm]
\hline \hline
$4$ & diag($1,1,1,0$) & diag($0,0,0,1$) & $-\frac{\pi}{2}\,<\,\alpha\,<\,\frac{\pi}{2}$ & ISO($3$)\\[1mm]
\hline
$5$ & diag($1,1,-1,0$) & diag($0,0,0,1$) & $-\frac{\pi}{2}\,<\,\alpha\,<\,\frac{\pi}{2}$ & ISO($2,1$)\\[1mm]
\hline \hline
$6$ & diag($1,1,0,0$) & diag($0,0,1,1$) & $-\frac{\pi}{4}\,<\,\alpha\,\le\,\frac{\pi}{4}$ & $\left\{\begin{array}{cc}\textrm{CSO}(2,0,2)\ , & \alpha\,\ne\,\frac{\pi}{4}\ ,\\ \mathfrak{f}_{1}\quad(\textrm{Solv}_{6}) \ , & \alpha\,=\,\frac{\pi}{4}\ .\end{array}\right.$\\[4mm]
\hline
$7$ & diag($1,1,0,0$) & diag($0,0,1,-1$) & $-\frac{\pi}{2}\,<\,\alpha\,<\,\frac{\pi}{2}$ & $\left\{\begin{array}{cc}\textrm{CSO}(2,0,2)\ , & |\alpha|\,<\,\frac{\pi}{4}\ ,\\ \textrm{CSO}(1,1,2)\ , & |\alpha|\,>\,\frac{\pi}{4}\ ,\\ \mathfrak{g}_{0}\quad(\textrm{Solv}_{6}) \ , & |\alpha|\,=\,\frac{\pi}{4}\ .\end{array}\right.$\\[4mm]
\hline
$8$ & diag($1,1,0,0$) & diag($0,0,0,1$) & $-\frac{\pi}{2}\,<\,\alpha\,<\,\frac{\pi}{2}$ & $\mathfrak{h}_{1}\quad(\textrm{Solv}_{6})$\\[1mm]
\hline
$9$ & diag($1,-1,0,0$) & diag($0,0,1,-1$) & $-\frac{\pi}{4}\,<\,\alpha\,\le\,\frac{\pi}{4}$ & $\left\{\begin{array}{cc}\textrm{CSO}(1,1,2)\ , & \alpha\,\ne\,\frac{\pi}{4}\ ,\\ \mathfrak{f}_{2}\quad(\textrm{Solv}_{6}) \ , & \alpha\,=\,\frac{\pi}{4}\ .\end{array}\right.$\\[4mm]
\hline
$10$ & diag($1,-1,0,0$) & diag($0,0,0,1$) & $-\frac{\pi}{2}\,<\,\alpha\,<\,\frac{\pi}{2}$ & $\mathfrak{h}_{2}\quad(\textrm{Solv}_{6})$\\[1mm]
\hline \hline $11$ & diag($1,0,0,0$) & diag($0,0,0,1$) &
$-\frac{\pi}{4}\,<\,\alpha\,\le\,\frac{\pi}{4}$ &
$\left\{\begin{array}{cc}\mathfrak{l}\quad(\textrm{Nil}_{6}(3)\,)\ , & \alpha\,\ne\,0\ ,\\
\textrm{CSO}(1,0,3)\ , &
\alpha\,=\,0\ .\end{array}\right.$\\[4mm]
\hline
\end{tabular}
}
\end{center}
{\it \caption{All the T-duality orbits of consistent gaugings in
half-maximal supergravity in $D=7$. Any value of $\,\alpha\,$
parameterises inequivalent orbits. More details about the
non-semisimple gauge algebras $\mathfrak{f}_{1}$,
$\mathfrak{f}_{2}$, $\mathfrak{h}_{1}$, $\mathfrak{h}_{2}$,
$\mathfrak{g}_{0}$ and $\mathfrak{l}$ are given in
appendix~\ref{appendix_A}.} \label{orbits_halfmax7}}
\end{table}

As we will see later, some of these consistent gaugings in general
include non-zero non-geometric fluxes, but at least in some of these
cases one will be able to dualise the given configuration to a
perfectly geometric background.

\subsubsection*{Higher-dimensional geometric origin}

Ten-dimensional heterotic string theory compactified on a $T^3$
gives rise to a half-maximal supergravity in $D=7$ where the
SL($4$)$\,=\,$SO($3,3$) factor in the global symmetry of this theory
can be interpreted as the T-duality group. The set of generalised
fluxes which can be turned on here is given by
\be \left\{f_{abc},\,{f_{ab}}^c,\,{f_a}^{bc},\,f^{abc}\right\}\equiv \left\{H_{abc},\,{\omega_{ab}}^c,\,{Q_a}^{bc},\,R^{abc}\right\}\
, \label{Fluxes}\ee
where $a,b,c\,=\,1,2,3$.

These are exactly the objects that one obtains by decomposing a
three-form of SO($3,3$) with respect to its GL($3$) subgroup. The
number of independent components of the above fluxes (including
traces of $\omega$ and $Q$) amounts to $1+9+9+1\,=\,20$, which is
the number of independent components of a three-form of SO($3,3$).
Nevertheless, the three-form representation is not irreducible since
the Hodge duality operator in 3+3 dimensions squares to 1. This
implies that one can always decompose it in a self-dual (SD) and
anti-self-dual (ASD) part
\be \textbf{10}\,\oplus\,\textbf{10}^\prime\quad\textrm{of
SL}(4)\quad\longleftrightarrow\quad\textbf{10}_{\textrm{SD}}\,\oplus\,\textbf{10}_{\textrm{ASD}}\quad\textrm{of
SO}(3,3)\ , \ee
such that the matching between the embedding tensor deformations
$(M_{mn},\,\tilde{M}^{mn})$ and the generalised fluxes given in
\eqref{Fluxes} now perfectly works. The explicit mapping between
vectors of SO($3,3$) expressed in light-cone coordinates and
two-forms of SL($4$) can be worked out by means of the SO($3,3$) 't
Hooft symbols $\left(G_A\right)^{mn}$ (see
Appendix~\ref{appendix_B}). This gives rise to the following
dictionary between the $M$ and $\tilde{M}$-components and the fluxes
given in \eqref{Fluxes}
\be
M\,=\,\textrm{diag}\,\left(H_{123},\,{Q_1}^{23},\,{Q_2}^{31},\,{Q_3}^{12}\right)\
,\quad
\tilde{M}\,=\,\textrm{diag}\,\left(R^{123},\,{\omega_{23}}^1,\,{\omega_{31}}^2,\,{\omega_{12}}^3\right)\
.\label{dictionary}\ee

The QC given in equations \eqref{QC1}-\eqref{QC4} enjoy a symmetry
in the exchange
\be (M,\,\xi)\,\overset{\eta}{\leftrightarrow}\,(-\tilde{M},\,-\xi)\
. \label{triple_duality}\ee
The discrete $\mathbb{Z}_2$ transformation $\eta$ corresponds to the
following O($3,3$) element with determinant $-1$
\be \eta\,=\,\left(
\begin{array}{cc}
0 & \mathds{1}_3\\
\mathds{1}_3 & 0
\end{array}\right)\ ,
\label{eta}\ee
which can be interpreted as a triple T-duality exchanging the three
compact coordinates $y^{a}$ with the corresponding winding
coordinates $\tilde{y}_{a}$ in the language of DFT.

Now we have all the elements to analyze the higher dimensional origin
of the orbits classified in table~\ref{orbits_halfmax7}.

\begin{itemize}

\item \textbf{Orbits 1 -- 3:}  These gaugings are non-geometric for
every $\alpha\ne 0$; for $\alpha =0$, they correspond to coset
reductions of heterotic string theory. See \emph{e.g.} the $S^{3}$
compactification in ref.~\cite{Cvetic:2000dm} giving rise to the
SO($4$) gauging. This theory was previously obtained in
ref.~\cite{Salam:1983fa} as $\mathcal{N}=2$ truncation of a maximal
supergravity in $D=7$.

\item \textbf{Orbits 4 -- 5:} For any value of $\alpha$ we can
always dualise these representatives to the one obtained by means of a
twisted $T^{3}$ reduction with $H$ and $\omega$ fluxes.

\item \textbf{Orbits 6 -- 7:} For any $\alpha\ne 0$ these orbits could be obtained from supergravity compactifications on locally-geometric T-folds, whereas for $\alpha=0$ it falls again in a special
case of the reductions described for orbits 4 and 5.

\item \textbf{Orbits 8 -- 11:} For any value of $\alpha$, these
orbits always contain a geometric representative involving less
general $H$ and $\omega$ fluxes.

\end{itemize}

To summarise, in the half-maximal $D=7$ case, we encounter a number
of orbits which do not have an obvious higher-dimensional origin. To
be more precise, these are orbits 1, 2 and 3 for $\alpha\ne 0$. The
challenge in the next subsection will be to establish what DFT can
do for us in order to give these orbits a higher-dimensional origin.
Again, before reading the following subsections we refer to the
section~\ref{subsec:twist_matrices} for a discussion of what we mean
by light-cone and cartesian formulations.

\subsubsection*{Higher-dimensional DFT origin}

First of all we would like to show here how to capture the gaugings that only
involve (up to duality rotations) fluxes $H_{abc}$ and
${\omega_{ab}}^{c}$. For this, we start from the light-cone
formulation, and propose the following Ansatz for a {\it globally geometric twist} (involving $e$ and $B$ and physical coordinates $y$)
\bea
e &=& \begin{pmatrix}1 & 0& \frac{\omega_1}{\omega_3} \sin (\omega_1\,\omega_3\,y^2) \\ 0 & \cos(\omega_2\,\omega_3\,y^1) & -\frac{\omega_2}{\omega_3} \cos (\omega_1\,\omega_3\,y^2) \sin(\omega_2\,\omega_3\,y^1) \\ 0 & \frac{\omega_3}{\omega_2} \sin(\omega_2\,\omega_3\,y^1) & \cos (\omega_1\,\omega_3\,y^2) \cos(\omega_2\,\omega_3\,y^1)\end{pmatrix}\ ,\\
B&=& \begin{pmatrix}0& 0 & 0\\ 0& 0& H\,y^1\,\cos
(\omega_1\,\omega_3\,y^2) \\ 0 & -H\,y^1\,\cos
(\omega_1\,\omega_3\,y^2)& 0\end{pmatrix} \ , \\
\lambda &=& - \frac 12 \log(\cos (\omega_1\omega_3 y^2)) \ .\eea
This is far from being the most general ansatz, but it serves our
purposes of reaching a large family of geometric orbits. The
parameters $\omega_i$ can be real, vanishing or imaginary, since $U$
is real and well-behaved in these cases. The QC, weak and strong
constraints are all automatically satisfied, and the gaugings read
\be M = {\rm diag} (H\ ,\ 0\ ,\ 0\ ,\ 0) \ ,\ \ \ \ \tilde M = {\rm
diag} (0\ , \ \omega_1^2\ ,\ \omega_2^2\ ,\ \omega_3^2)\ . \ee From
here, by choosing appropriate values of the parameters the {\bf
orbits 4, 5, 8, 10} and {\bf 11} can be obtained. Indeed these are
geometric as they only involve gauge and (geo)metric fluxes.

Secondly, in order to address the remaining orbits, we consider an SO($2,2$) twist $U_{4}$ embedded in O($3,3$) in the
following way
\be
U =
\begin{pmatrix}1 & 0 & 0 & 0\\ 0& A & 0 & B\\ 0& 0 & 1 & 0 \\ 0& C &
0 & D
\end{pmatrix}\ , \ \ \ \ \ \ U_{4} = \begin{pmatrix}A& B \\ C &
D\end{pmatrix}\ ,  \ \ \ \ \ \ \lambda =0\ . \ee
This situation is analog to the SO$(1,1)$ twist considered in the $D
= 8$ case, but with a more general twist. Working in the cartesian formulation, one can define the generators
and elements of SO$(2,2)$ as \be [t_{IJ}]_K{}^L = \delta^L_{[I}
\eta_{J]K}\ , \ \ \ \ \  U_4 = \exp\left(t_{IJ} \phi^{IJ}\right)\ ,
\ee where the rotations are generated by $t_{12}$ and $t_{34}$, and
the boosts by the other generators. Also, we take $\phi^{IJ} =
\alpha^{IJ} y^1 + \beta^{IJ} \tilde y_1$ to be linear.

From the above $\textrm{SO}(2,2)$ duality element one can reproduce the following orbits employing a {\it locally geometric twist} (including $e$, $B$ and $\beta$ but only depending on $y$, usually referred to as a T-fold):
~
\begin{itemize}

\item {\bf Orbit 6}  can be obtained by taking
\be {\bf (6)} \ \ \ \alpha^{12} = - \beta^{12} = - \frac
1{\sqrt{2}}\,(\cos \alpha + \sin \alpha)   \ , \ \ \ \alpha^{34} =
-\beta^{34}= - \frac 1{\sqrt{2}}\,(\cos \alpha + \sin \alpha)\
.\nn\ee
and all other vanishing.

\item \textbf{Orbits 7} and {\textbf 9}  can be obtained by the following particular identifications
\be
\begin{array}{lclclc}
\phi^{14} = \phi^{23} & , & \phi^{12} = \phi^{34} & \textrm{and} &
\phi^{13} = \phi^{24} & .\end{array}\nn\ee
\be {\bf (7)} \ \ \ \alpha^{14} = - \beta^{14} = -
\frac{1}{\sqrt{2}}\,\sin \alpha \ , \ \ \ \alpha^{12} = - \beta^{12}
= - \frac{1}{\sqrt{2}}\,\cos \alpha \ , \ \ \ \alpha^{13} =
\beta^{13} =0 \ ,\nn\ee
\be {\bf (9)} \ \ \ \alpha^{14} = - \beta^{14} = -
\frac{1}{\sqrt{2}}\,\sin \alpha \ , \ \ \ \alpha^{12} = \beta^{12} =
0 \ , \ \ \ \alpha^{13} = \beta^{13} = - \frac{1}{\sqrt{2}}\,\cos
\alpha \ .\nn\ee
\end{itemize}
All these backgrounds satisfy both the weak and the strong
constraints  and hence they admit a locally geometric
description. This is in agreement with the fact that the simplest
representative of \textbf{orbits 6, 7} and {\textbf 9} given in
table~\ref{orbits_halfmax7} contains $H$, $\omega$ and $Q$ fluxes
but no $R$ flux.

Finally, one can employ the same SO($2,2$) duality elements with
different identifications to generate the remaining orbits with a
{\it non-geometric twist} (involving both $y$ and $\tilde y$
coordinates): ~
\begin{itemize}

\item \textbf{Orbits 1, 3}  can be again obtained by considering an
SO($2)\,\times\,$SO($2$) twist  with arbitrary $\phi^{12}$ and
$\phi^{34}$:
\be {\bf (1)} \ \ \ \alpha^{12} = - 2\,\sqrt{2}\,(\cos \alpha + \sin
\alpha) \ , \ \ \ \beta^{34} = 2\,\sqrt{2}\,(\cos \alpha - \sin
\alpha)\ , \ \ \ \alpha^{34} = \beta^{12} = 0 \ ,\nn\ee
\be {\bf (3)} \ \ \ \alpha^{34} = - 2\,\sqrt{2}\,(\cos \alpha + \sin
\alpha) \ , \ \ \ \beta^{12} = 2\,\sqrt{2}\,(\cos \alpha - \sin
\alpha)\ , \ \ \ \alpha^{12} = \beta^{34} = 0 \ .\nn\ee
\item \textbf{Orbit 2} can be obtained by means of a different
SO($2,2$) twist built out of the two rotations and two boosts
subject to the following identification
\be
\begin{array}{lcl}
\phi^{14} = \phi^{23} & \textrm{, } & \phi^{12} = \phi^{34}\ .
\end{array}
\ee
\be {\bf (2)} \ \ \ \alpha^{14} = \beta^{12} = \frac
1{\sqrt{2}}\,(\cos \alpha - \sin \alpha) \ , \ \ \ \alpha^{12} = -
\beta^{14} = - \frac 1{\sqrt{2}}\,(\cos \alpha + \sin \alpha)\ .
\nn\ee
\end{itemize}
These backgrounds violate both the weak and the strong constraints for $\alpha \neq 0$.
This implies that these backgrounds are truly doubled and they do
not even admit a locally geometric description.

Finally, let us also give an example of degeneracy in twist orbits-space reproducing the same orbit of gaugings. The following twist
\be
\phi^{12} = \phi^{13} \ , \ \ \ \phi^{34} = \phi^{24} \ , \ \ \ \phi^{23} = \phi^{14} = 0
\ee
\be {\bf (6)} \ \ \ \alpha^{13} = - \frac{1}{\sqrt{2}}(\cos \alpha + \sin
\alpha) \ , \ \ \ \beta^{24} = \frac{1}{\sqrt{2}}(\cos \alpha - \sin
\alpha)\ , \ \ \ \alpha^{24} = \beta^{13} = 0 \ ,\nn\ee
also reproduces the {\bf orbit 6}, but in this case through a non-geometric twist. What happens in this case is that although the twist matrix does not satisfy the weak/strong constraints, the contractions in (\ref{Max_VS_Geom}) cancel.
\section{Conclusions}
\label{sec:conclusions}

In this paper we have provided a litmus test to the notion of  non-geometry, by classifying the explicit orbits of consistent gaugings of different supergravity theories, and considering the possible higher-dimensional origins of these. The results turn out to be fundamentally different for the cases of U-duality orbits of maximal supergravities, and T-duality orbits of half-maximal theories.

In the former case we have managed to explicitly classify all U-duality orbits in dimensions $8 \leq D \leq 11$. This led to zero, one, four and ten discrete orbits in dimensions $D=11, 10,9$ and $8$, respectively, with different associated gauge groups. Remarkably, we have found that all of these orbits have a higher-dimensional origin via some geometric compactification, be it twisted reductions or compactifications on group manifolds or coset spaces. In our parlance, we have therefore found that all U-duality orbits are geometric. The structure of U-duality orbits is therefore dramatically different from the sketch of figure 1 in the introduction. Although a full classification of all orbits in lower-dimensional cases becomes increasingly cumbersome, we are not aware of any examples that are known to be non-geometric. It could therefore hold in full generality that all U-duality orbits are necessarily geometric.

This is certainly not the case for T-duality orbits of gaugings of half-maximal supergravities. In this case, we have provided the explicit classification in dimensions $7 \leq D \leq 10$ (where in $D=7$ we have only included three-form fluxes). The numbers of distinct families of orbits in this case are zero, one, three and eleven in dimensions $D=10,9,8$ and $7$, respectively, which includes both discrete and one-parameter orbits. A number of these orbits do not have a higher-dimensional origin in terms of a geometric compactification. Such cases are {\bf orbits 2} and {\bf 3} in $D=8$ and {\bf orbits 1, 2} and {\bf 3} in $D=7$ for $\alpha\neq 0$. Indeed, these are exactly the orbits that do not admit an uplift to the maximal theory. As proven in section~\ref{subsec:GDFT}, all such orbits necessarily violate the weak and/or strong constraints, and therefore need truly doubled backgrounds. Thus, the structure of T-duality orbits is very reminiscent of figure 1 in the introduction. Given the complications that already arise in these simpler higher-dimensional variants, one can anticipate that the situation will be similar in four-dimensional half-maximal supergravity.

Fortunately, the formalism of double field theory seems tailor-made to
generate additional T-duality orbits of half-maximal supergravity. Building on the recent generalisation of the definition of
double field theory \cite{Grana:2012rr}, we have demonstrated that
all T-duality orbits, including the non-geometric ones in $D=7,8$,
can be generated by a twisted reduction of double field theory. We
have explicitly provided duality twists for all orbits. For locally-geometric orbits the twists only depend on the physical
coordinates $y$, while for the non-geometric orbits these
necessarily also include $\tilde y$. Again, based on our exhaustive
analysis in higher-dimensions, one could conjecture that also in
lower-dimensional theories, all T-duality orbits follow from this
generalised notion of double field theory.

At this point we would like to stress once more that a given orbit of gaugings can be generated from different twist orbits. Therefore, there is a degeneracy in the space of twist orbits giving rise to a particular orbit of gaugings. Interestingly, as it is the case of {\bf orbit 6} in $D=7$  for instance, one might find two different twist orbits reproducing the same orbit of gaugings, one  violating weak and strong constraints, the other one satisfying both. Our notion of a locally
geometric orbit of gaugings is related to the existence of at least one undoubled background giving rise to it. However, this ambiguity seems to be peculiar of gaugings containing $Q$ flux. These can, in principle, be independently obtained by
either adding a $\beta$ but no $\tilde{y}$ dependence (locally geometric choice, usually called T-fold), or by including non-trivial $\tilde{y}$
dependence but no $\beta$ (non-geometric choice) \cite{Aldazabal:2011nj}.

Another remarkable degeneracy occurs for the case of semi-simple
gaugings, corresponding to {\bf orbits 1 -- 3} in $D=7$. For the
special case of $\alpha = 0$, we have two possible ways of
generating such orbits from higher-dimensions: either a coset
reduction over a sphere or analytic continuations thereof, or a
duality twist involving non-geometric coordinate dependence.
Therefore $d$-dimensional coset reductions seem to be equivalent to
$2d$-dimensional twisted torus reductions (with the latter in fact
being more general, as it leads to all values of $\alpha$).
Considering the complications that generally arise in proving the
consistency of coset reductions, this is a remarkable reformulation
that would be interesting to understand in more detail. Furthermore,
when extending the notion of double field theory to type II and
M-theory, this relation could also shed new light on the consistency
of the notoriously difficult four-, five- and seven-sphere
reductions of these theories.

Our results mainly focus on Scherk-Scharz compactifications leading to gauged supergravities with vanishing $\xi_M$ fluxes. In addition, we have restricted to the NSNS sector and ignored $\alpha'$-effects.
Also, we stress once again that relaxing the strong and weak constraints
is crucial in part of our analysis. If we kept the weak constraint, typically the Jacobi identities would lead to backgrounds satisfying also the strong constraint\footnote{We thank O.~Hohm for enlightening correspondence on this point.} \cite{Grana:2012rr}.
However, from a purely (double) field theoretical analysis the weak constraint is not necessary. A sigma model
analysis beyond tori would help us to clarify the relation between DFT without
the weak and strong constraints and string field theory on more
general backgrounds. We hope to come back to this point in the
future.

\section*{Acknowledgements}

We thank G.~Aldazabal, D.~Andriot, G.~Dall'Agata, M.~Gra$\tilde{\textrm{n}}$a, A.~Guarino, O.~Hohm, M.~Larfors, C.~Nu\~nez and H.~Samtleben for interesting and stimulating
discussions. We are also grateful to G.~Aldazabal, M.~Gra$\tilde{\textrm{n}}$a, O.~Hohm and C.~Nu\~nez for valuable comments on a draft version of this manuscript. J.J.F.M. would like to express his gratitude to the CTN
Groningen for its warm hospitality while part of this project was
done. The work of G.D. and D.R. is supported by a VIDI grant from
the Netherlands Organisation for Scientific Research (NWO). The work
of D.M. is supported by the ERC Starting Independent Researcher
Grant 259133 - ObservableString. The work of J.J.F.M. has been supported by the Spanish Ministry of Education FPU grant AP2008-00919.

\appendix

\section{Different solvable and nilpotent gaugings}\label{appendix_A}

In section~\ref{subsec:Half_Max7} we have studied the T-duality
orbits of gaugings in half-maximal $D=7$ supergravity and for each
of them, we identified the gauge algebra and presented the results in
table~\ref{orbits_halfmax7}. Since there is no exhaustive
classification of non-semisimple algebras of dimension 6, we would
like to explicitly give the form of the algebras appearing in
table~\ref{orbits_halfmax7}.

\subsection*{Solvable algebras}

\subsubsection*{The CSO($2,0,2$) and CSO($1,1,2$) algebras}

The details about these algebras can be found in
ref.~\cite{deRoo:2006ms}; we summarise here some relevant facts.

The six generators are labelled as
$\{t_{0},\,t_{i},\,s_{i},\,z\}_{i=1,2}$, where $t_{0}$ generates
SO($2$) (SO($1,1$)), under which $\{t_{i}\}$ and $\{s_{i}\}$
transform as doublets
\be
\begin{array}{cccc}
\left[t_{0},\,t_{i}\right]\,=\,{\epsilon_{i}}^{j}\,t_{j} & , &
\left[t_{0},\,s_{i}\right]\,=\,{\epsilon_{i}}^{j}\,s_{j} & ,
\end{array}
\ee
where the Levi-Civita symbol ${\epsilon_{i}}^{j}$ has one index
lowered with the metric $\eta_{ij}\,=\,$diag$(\pm 1,1)$ depending on
the two different signatures. $z$ is a central charge appearing in
the following commutators
\be \left[t_{i},\,s_{j}\right]\,=\,\delta_{ij}\,z\ . \ee
The Cartan-Killing metric is diag($\mp 1, \underbrace{0, \cdots,
0}_{\textrm{6 times}}$), where the $\mp$ is again related to the two
different signatures.

\subsubsection*{The $\mathfrak{f}_{1}$ and $\mathfrak{f}_{2}$ algebras}

These are of the form Solv$_{4}\,\times\,$U$(1)^{2}$. The 4
generators of Solv$_{4}$ are labeled by
$\{t_{0},\,t_{i},\,z\}_{i=1,2}$, where $t_{0}$ generates SO($2$)
(SO($1,1$)), under which $\{t_{i}\}$ transform as a doublet
\be \left[t_{0},\,t_{i}\right]\,=\,{\epsilon_{i}}^{j}\,t_{j}\ , \ee
\be \left[t_{i},\,t_{j}\right]\,=\,\epsilon_{ij}\,z\ . \ee
The Cartan-Killing metric is diag($\mp 1, \underbrace{0, \cdots,
0}_{\textrm{6 times}}$).

\subsubsection*{The $\mathfrak{h}_{1}$ and $\mathfrak{h}_{2}$ algebras}

The 6 generators are $\{t_{0},\,t_{i},\,s_{i},\,z\}_{i=1,2}$ and
they satisfy the following commutation relations
\be
\begin{array}{lclc}
\left[t_{0},\,t_{i}\right]\,=\,{\epsilon_{i}}^{j}\,t_{j} & , &
\left[t_{0},\,s_{i}\right]\,=\,{\epsilon_{i}}^{j}\,s_{j}\,+\,t_{i} &
, \\[2mm]
\left[t_{i},\,s_{j}\right]\,=\,\delta_{ij}\,z & , &
\left[s_{i},\,s_{j}\right]\,=\,\epsilon_{ij}\,z & .
\end{array}
\ee
The Cartan-Killing metric is diag($\mp 1, \underbrace{0, \cdots,
0}_{\textrm{6 times}}$).

\subsubsection*{The $\mathfrak{g}_{0}$ algebra}

The 6 generators are $\{t_{0},\,t_{I},\,z\}_{I=1,\cdots,4}$, where
$t_{0}$ transforms cyclically the $\{t_{I}\}$ amongst themselves
such that
\be
\left[\left[\left[\left[t_{I},\,t_{0}\right],\,t_{0}\right],\,t_{0}\right],\,t_{0}\right]\,=\,t_{I}\
, \ee
and
\be \left[t_{1},\,t_{3}\right]\,=\,\left[t_{2},\,t_{4}\right]\,=\,z\
.\ee
Note that this algebra is solvable and not nilpotent even though its
Cartan-Killing metric is \emph{completely zero}.

\subsection*{Nilpotent algebras}

\subsubsection*{The CSO($1,0,3$) algebra}

The details about this algebra can be again found in
ref.~\cite{deRoo:2006ms}; briefly summarizing, the 6 generators are
given by  $\{t_{m},\,z^{m}\}_{m=1,2,3}$ and they satisfy the
following commutation relations
\be \left[t_{m},\,t_{n}\right]\,=\,\epsilon_{mnp}\,z^{p}\ ,\ee
with all the other brackets being vanishing. The order of nilpotency
of this algebra is 2.

\subsubsection*{The $\mathfrak{l}$ algebra}

The 6 generators $\{t_{1},\cdots,\,t_{6}\}$ satisfy the following
commutation relations
\be
\begin{array}{lclclc}
 \left[t_{1},\,t_{2}\right]\,=\,t_{4} & , &
 \left[t_{1},\,t_{4}\right]\,=\,t_{5} & , &
 \left[t_{2},\,t_{4}\right]\,=\,t_{6} & .
\end{array}
\ee
The corresponding central series reads
\be
\begin{array}{lclclcl}
 \left\{t_{1},\,t_{2},\,t_{3},\,t_{4},\,t_{5},\,t_{6}\right\} &\supset&
 \left\{t_{4},\,t_{5},\,t_{6}\right\} &\supset&
 \left\{t_{5},\,t_{6}\right\} &\supset& \left\{0\right\}\ ,
\end{array}
\ee
from which we can immediately conclude that its nilpotency order is
3.

\section{SO($2,2$) and SO($3,3$) 't Hooft symbols}\label{appendix_B}

In section~\ref{sec:DFT} we discuss the origin of a given flux
configuration from DFT backgrounds specified by twist matrices $U$.
The deformations of half-maximal supergravity in $D=10-d$ which can
be interpreted as the gauging of a subgroup of the T-duality group
O($d,d$) can be described by a 3-form of O($d,d$) $f_{ABC}$ which
represents a certain (non-)geometric flux configuration.

In $D=8$ and $D=7$, the T-duality group happens to be isomorphic to
SL($2)\,\times\,$SL($2$) and SL($4$) respectively. As a consequence,
in order to explicitly relate flux configurations and embedding
tensor orbits, we need to construct the mapping between T-duality
irrep's and irrep's of SL($2)\,\times\,$SL($2$) and SL($4$)
respectively.

\subsection*{From the $(\textbf{2},\textbf{2})$ of SL($2)\,\times\,$SL($2$) to the $\textbf{4}$ of SO($2,2$)}

The 't Hooft symbols $\left(G_A\right)^{\alpha i}$ are invariant
tensors which map the fundamental representation of SO($2,2$) (here
denoted by $A$), into the $(\textbf{2},\textbf{2})$ of
SL($2)\,\times\,$SL($2$)
\be v^{\alpha i}\,=\,\left(G_A\right)^{\alpha i}\,v^{A}\,,
\label{vec2vecvec}\ee
where $v^{A}$ denotes a vector of SO($2,2$) and the indices $\alpha$
and $i$ are raised and lowered by means of $\epsilon_{\alpha\beta}$
and $\epsilon_{ij}$ respectively. $\left(G_A\right)^{\alpha i}$ and
$\left(G_A\right)_{\alpha i}$ satisfy the following identities
\bea
&&\left(G_A\right)_{\alpha i}\,\left(G_B\right)^{\alpha i}\,=\,\eta_{AB}\ ,\\
&&\left(G_A\right)^{\alpha i}\,\left(G^{A}\right)^{\beta
j}\,=\,\epsilon^{\alpha \beta}\,\epsilon^{ij}\ , \eea
where $\eta_{AB}$ is the SO($2,2$) metric.

After choosing light-cone coordinates for SO($2,2$), our choice for
the tensors $\left(G_A\right)^{\alpha i}$ is the following
\begin{align}
\left(G_1\right)^{\alpha i} &= \left(
  \begin{array}{cc}
  0 & 0 \\
  0 & 1
  \end{array}
  \right)\,\,,
& \left(G_2\right)^{\alpha i} &= \left(
  \begin{array}{cc}
  0 & 1 \\
  0 & 0
  \end{array}
  \right)\,\,,
\\
\left(G_{\bar{1}}\right)^{\alpha i} &= \left(
  \begin{array}{cc}
  1 & 0 \\
  0 & 0
  \end{array}
  \right)\,\,,
& \left(G_{\bar{2}}\right)^{\alpha i} &= \left(
  \begin{array}{cc}
  0 & 0 \\
  -1 & 0
  \end{array}
  \right)\,\,.
\end{align}
By making use of the mapping \eqref{vec2vecvec}, we can rewrite the
structure constants $(X_{\alpha i})_{\beta j}{}^{\gamma k}$ as a
3-form of SO($2,2$) as follows:
\begin{align}
 f_{ABC}
&= (X_{\alpha i})_{\beta j}{}^{\gamma k} (G_{A})^{\alpha i}
(G_{B})^{\beta j} (G_{C})_{\gamma k}\,\,. \label{X2f_D=8}
\end{align}

\subsection*{From the $\textbf{6}$ of SL($4$) to the $\textbf{6}$ of SO($3,3$)}

The 't Hooft symbols $\left(G_A\right)^{mn}$ are invariant tensors
which map the fundamental representation of SO($3,3$), \emph{i.e.}
the \textbf{6} into the anti-symmetric two-form of SL($4$)
\be v^{mn}\,=\,\left(G_A\right)^{mn}\,v^{A}\,,
\label{vec2two-form}\ee
where $v^{A}$ denotes a vector of SO($3,3$). The two-form irrep of
SL($4$) is real due to the role of the Levi-Civita tensor relating
$v^{mn}$ to $v_{mn}$
\be v_{mn}\,=\,\frac{1}{2}\,\epsilon_{mnpq}\,v^{pq}\,. \ee
The 't Hooft symbols with lower SL($4$) indices
$\left(G_A\right)_{mn}$ carry out the inverse mapping of the one
given in \eqref{vec2two-form}. The tensors $\left(G_A\right)^{mn}$
and
$\left(G_A\right)_{mn}=\frac{1}{2}\,\epsilon_{mnpq}\,\left(G_A\right)^{pq}$
satisfy the following identities
\bea
&&\left(G_A\right)_{mn}\,\left(G_B\right)^{mn}\,=\,2\,\eta_{AB}\,,\\
&&\left(G_{A}\right)_{mp}\,\left(G_{B}\right)^{pn}\,+\left(G_{B}\right)_{mp}\,\left(G_{A}\right)^{pn}\,=\,-\delta_m^{n}\,\eta_{AB}\,,\\
&&\left(G_A\right)_{mp}\,\left(G_B\right)^{pq}\,\left(G_C\right)_{qr}\,\left(G_D\right)^{rs}\,\left(G_E\right)_{st}\,\left(G_F\right)^{tn}\,=\,\delta_m^{n}\,\epsilon_{ABCDEF}\,,
\eea
where $\eta_{AB}$ and $\epsilon_{ABCDEF}$ are the SO($3,3$) metric
and Levi-Civita tensor respectively.

After choosing light-cone coordinates for SO($3,3$) vectors, our
choice of the 't Hooft symbols is
\begin{align}
\left(G_1\right)^{mn}
&=
\left(
  \begin{array}{cccc}
  0 & -1 & 0 & 0\\
  1 & 0 & 0 & 0\\
  0 & 0 & 0 & 0\\
  0 & 0 & 0 & 0
  \end{array}
  \right)\,\,,
&
\left(G_2\right)^{mn}
&=
\left(
  \begin{array}{cccc}
  0 & 0 & -1 & 0\\
  0 & 0 & 0 & 0\\
  1 & 0 & 0 & 0\\
  0 & 0 & 0 & 0
  \end{array}
  \right)\,\,,
\\
\left(G_3\right)^{mn}
&=
\left(
  \begin{array}{cccc}
  0 & 0 & 0 & -1\\
  0 & 0 & 0 & 0\\
  0 & 0 & 0 & 0\\
  1 & 0 & 0 & 0
  \end{array}
  \right)\,\,,
&
\left(G_{\bar1}\right)^{mn}
&=
\left(
  \begin{array}{cccc}
  0 & 0 & 0 & 0\\
  0 & 0 & 0 & 0\\
  0 & 0 & 0 & -1\\
  0 & 0 & 1 & 0
  \end{array}
  \right)\,\,,
\\
\left(G_{\bar2}\right)^{mn}
&=
\left(
  \begin{array}{cccc}
  0 & 0 & 0 & 0\\
  0 & 0 & 0 & -1\\
  0 & 0 & 0 & 0\\
  0 & 1 & 0 & 0
  \end{array}
  \right)\,\,,
&
\left(G_{\bar3}\right)^{mn}
&=
\left(
  \begin{array}{cccc}
  0 & 0 & 0 & 0\\
  0 & 0 & 1 & 0\\
  0 & -1 & 0 & 0\\
  0 & 0 & 0 & 0
  \end{array}
  \right)\,\,.
\end{align}

Thus, we can rewrite the structure constants in the \textbf{6},
$(X_{mn})_{pq}{}^{rs}$, arising from (\ref{gen_half-max}) as a
3-form of SO($3,3$) as follows:
\begin{align}
 f_{ABC}
&= (X_{mn})_{pq}{}^{rs} (G_{A})^{mn} (G_{B})^{pq} (G_{C})_{rs}\,\,.
\label{X2f_D=7}
\end{align}

%
%


\bibliography{references}

\providecommand{\href}[2]{#2}\begingroup\raggedright\begin{thebibliography}{10}

\bibitem{Schon:2006kz}
J.~Schon and M.~Weidner, ``{Gauged N=4 supergravities},''
  \href{http://dx.doi.org/10.1088/1126-6708/2006/05/034}{{\em JHEP} {\bf 0605}
  (2006)  034},
\href{http://arxiv.org/abs/hep-th/0602024}{{\tt arXiv:hep-th/0602024
  [hep-th]}}.

\bibitem{deWit:2007mt}
B.~de~Wit, H.~Samtleben, and M.~Trigiante, ``{The maximal D=4
  supergravities},''
  \href{http://dx.doi.org/10.1088/1126-6708/2007/06/049}{{\em JHEP} {\bf 06}
  (2007)  049}, \href{http://arxiv.org/abs/0705.2101}{{\tt arXiv:0705.2101
  [hep-th]}}.

\bibitem{Roest:2009dq}
D.~Roest, ``{Gaugings at angles from orientifold reductions},''
  \href{http://dx.doi.org/10.1088/0264-9381/26/13/135009}{{\em
  Class.Quant.Grav.} {\bf 26} (2009)  135009},
\href{http://arxiv.org/abs/0902.0479}{{\tt arXiv:0902.0479 [hep-th]}}.

\bibitem{Dall'Agata:2009gv}
G.~Dall'Agata, G.~Villadoro, and F.~Zwirner, ``{Type-IIA flux compactifications
  and N=4 gauged supergravities},''
  \href{http://dx.doi.org/10.1088/1126-6708/2009/08/018}{{\em JHEP} {\bf 08}
  (2009)  018}, \href{http://arxiv.org/abs/0906.0370}{{\tt arXiv:0906.0370
  [hep-th]}}.

\bibitem{Shelton:2005cf}
J.~Shelton, W.~Taylor, and B.~Wecht, ``{Nongeometric flux compactifications},''
  \href{http://dx.doi.org/10.1088/1126-6708/2005/10/085}{{\em JHEP} {\bf 0510}
  (2005)  085},
\href{http://arxiv.org/abs/hep-th/0508133}{{\tt arXiv:hep-th/0508133
  [hep-th]}}.

\bibitem{Hull:2009mi}
C.~Hull and B.~Zwiebach, ``{Double Field Theory},''
  \href{http://dx.doi.org/10.1088/1126-6708/2009/09/099}{{\em JHEP} {\bf 0909}
  (2009)  099},
\href{http://arxiv.org/abs/0904.4664}{{\tt arXiv:0904.4664 [hep-th]}}.

\bibitem{Aldazabal:2011nj}
G.~Aldazabal, W.~Baron, D.~Marques, and C.~Nunez, ``{The effective action of
  Double Field Theory},'' \href{http://dx.doi.org/10.1007/JHEP11(2011)052,
  10.1007/JHEP11(2011)109, 10.1007/JHEP11(2011)052,
  10.1007/JHEP11(2011)109}{{\em JHEP} {\bf 1111} (2011)  052},
  \href{http://arxiv.org/abs/1109.0290}{{\tt arXiv:1109.0290 [hep-th]}}.

\bibitem{Geissbuhler:2011mx}
D.~Geissbuhler, ``{Double Field Theory and N=4 Gauged Supergravity},''
  \href{http://dx.doi.org/10.1007/JHEP11(2011)116}{{\em JHEP} {\bf 1111} (2011)
   116}, \href{http://arxiv.org/abs/1109.4280}{{\tt arXiv:1109.4280 [hep-th]}}.

\bibitem{Grana:2012rr}
M.~Grana and D.~Marques, ``{Gauged Double Field Theory},''
\href{http://arxiv.org/abs/1201.2924}{{\tt arXiv:1201.2924 [hep-th]}}.

\bibitem{Scherk:1979zr}
J.~Scherk and J.~H. Schwarz, ``{How to Get Masses from Extra Dimensions},''
{\em Nucl.Phys.} {\bf B153} (1979)  61--88.

\bibitem{Hull:2009zb}
C.~Hull and B.~Zwiebach, ``{The Gauge algebra of double field theory and
  Courant brackets},''
  \href{http://dx.doi.org/10.1088/1126-6708/2009/09/090}{{\em JHEP} {\bf 0909}
  (2009)  090},
\href{http://arxiv.org/abs/0908.1792}{{\tt arXiv:0908.1792 [hep-th]}}.

\bibitem{Hohm:2010jy}
O.~Hohm, C.~Hull, and B.~Zwiebach, ``{Background independent action for double
  field theory},'' \href{http://dx.doi.org/10.1007/JHEP07(2010)016}{{\em JHEP}
  {\bf 1007} (2010)  016},
\href{http://arxiv.org/abs/1003.5027}{{\tt arXiv:1003.5027 [hep-th]}}.

O.~Hohm, C.~Hull, and B.~Zwiebach, ``{Generalized metric formulation of double
  field theory},'' \href{http://dx.doi.org/10.1007/JHEP08(2010)008}{{\em JHEP}
  {\bf 1008} (2010)  008},
\href{http://arxiv.org/abs/1006.4823}{{\tt arXiv:1006.4823 [hep-th]}}.

\bibitem{Hull:2004in}
C.~Hull, ``{A Geometry for non-geometric string backgrounds},''
  \href{http://dx.doi.org/10.1088/1126-6708/2005/10/065}{{\em JHEP} {\bf 0510}
  (2005)  065},
\href{http://arxiv.org/abs/hep-th/0406102}{{\tt arXiv:hep-th/0406102
  [hep-th]}}.

A.~Dabholkar and C.~Hull, ``{Generalised T-duality and non-geometric
  backgrounds},'' \href{http://dx.doi.org/10.1088/1126-6708/2006/05/009}{{\em
  JHEP} {\bf 0605} (2006)  009},
\href{http://arxiv.org/abs/hep-th/0512005}{{\tt arXiv:hep-th/0512005
  [hep-th]}}.

\bibitem{Andriot:2011uh}
D.~Andriot, M.~Larfors, D.~Lust, and P.~Patalong, ``{A ten-dimensional action
  for non-geometric fluxes},''
  \href{http://dx.doi.org/10.1007/JHEP09(2011)134}{{\em JHEP} {\bf 1109} (2011)
   134},
\href{http://arxiv.org/abs/1106.4015}{{\tt arXiv:1106.4015 [hep-th]}}.

D.~Andriot, O.~Hohm, M.~Larfors, D.~Lust, and P.~Patalong, ``{A geometric
  action for non-geometric fluxes},''
\href{http://arxiv.org/abs/1202.3060}{{\tt arXiv:1202.3060 [hep-th]}}.

\bibitem{Hohm:2011ex}
O.~Hohm and S.~K. Kwak, ``{Double Field Theory Formulation of Heterotic
  Strings},'' \href{http://dx.doi.org/10.1007/JHEP06(2011)096}{{\em JHEP} {\bf
  1106} (2011)  096},
\href{http://arxiv.org/abs/1103.2136}{{\tt arXiv:1103.2136 [hep-th]}}.

O.~Hohm and S.~K. Kwak, ``{Frame-like Geometry of Double Field Theory},''
  \href{http://dx.doi.org/10.1088/1751-8113/44/8/085404}{{\em J.Phys.A} {\bf
  A44} (2011)  085404},
\href{http://arxiv.org/abs/1011.4101}{{\tt arXiv:1011.4101 [hep-th]}}.

\bibitem{Jeon:2010rw}
I.~Jeon, K.~Lee, and J.-H. Park, ``{Differential geometry with a projection:
  Application to double field theory},''
  \href{http://dx.doi.org/10.1007/JHEP04(2011)014}{{\em JHEP} {\bf 1104} (2011)
   014},
\href{http://arxiv.org/abs/1011.1324}{{\tt arXiv:1011.1324 [hep-th]}}.

D.~S. Berman and M.~J. Perry, ``{Generalized Geometry and M theory},''
  \href{http://dx.doi.org/10.1007/JHEP06(2011)074}{{\em JHEP} {\bf 1106} (2011)
   074},
\href{http://arxiv.org/abs/1008.1763}{{\tt arXiv:1008.1763 [hep-th]}}.

D.~S. Berman, H.~Godazgar, M.~J. Perry, and P.~West, ``{Duality Invariant
  Actions and Generalised Geometry},''
  \href{http://dx.doi.org/10.1007/JHEP02(2012)108}{{\em JHEP} {\bf 1202} (2012)
   108},
\href{http://arxiv.org/abs/1111.0459}{{\tt arXiv:1111.0459 [hep-th]}}.

D.~S. Berman, H.~Godazgar, and M.~J. Perry, ``{SO(5,5) duality in M-theory and
  generalized geometry},''
  \href{http://dx.doi.org/10.1016/j.physletb.2011.04.046}{{\em Phys.Lett.} {\bf
  B700} (2011)  65--67},
\href{http://arxiv.org/abs/1103.5733}{{\tt arXiv:1103.5733 [hep-th]}}.

A.~Coimbra, C.~Strickland-Constable, and D.~Waldram, ``{$E_{d(d)} \times
  \mathbb{R}^+$ Generalised Geometry, Connections and M Theory},''
\href{http://arxiv.org/abs/1112.3989}{{\tt arXiv:1112.3989 [hep-th]}}.

A.~Coimbra, C.~Strickland-Constable, and D.~Waldram, ``{Supergravity as
  Generalised Geometry I: Type II Theories},''
  \href{http://dx.doi.org/10.1007/JHEP11(2011)091}{{\em JHEP} {\bf 1111} (2011)
   091},
\href{http://arxiv.org/abs/1107.1733}{{\tt arXiv:1107.1733 [hep-th]}}.

O.~Hohm and S.~K. Kwak, ``{N=1 Supersymmetric Double Field Theory},''
\href{http://arxiv.org/abs/1111.7293}{{\tt arXiv:1111.7293 [hep-th]}}.

I.~Jeon, K.~Lee, and J.-H. Park, ``{Supersymmetric Double Field Theory: Stringy
  Reformulation of Supergravity},''
\href{http://arxiv.org/abs/1112.0069}{{\tt arXiv:1112.0069 [hep-th]}}.

N.~B. Copland, ``{A Double Sigma Model for Double Field Theory},''
\href{http://arxiv.org/abs/1111.1828}{{\tt arXiv:1111.1828 [hep-th]}}.

D.~S. Berman, E.~T. Musaev, and M.~J. Perry, ``{Boundary Terms in Generalized
  Geometry and doubled field theory},''
  \href{http://dx.doi.org/10.1016/j.physletb.2011.11.019}{{\em Phys.Lett.} {\bf
  B706} (2011)  228--231},
\href{http://arxiv.org/abs/1110.3097}{{\tt arXiv:1110.3097 [hep-th]}}.

I.~Jeon, K.~Lee, and J.-H. Park, ``{Incorporation of fermions into double field
  theory},'' \href{http://dx.doi.org/10.1007/JHEP11(2011)025}{{\em JHEP} {\bf
  1111} (2011)  025},
\href{http://arxiv.org/abs/1109.2035}{{\tt arXiv:1109.2035 [hep-th]}}.

I.~Jeon, K.~Lee, and J.-H. Park, ``{Stringy differential geometry, beyond
  Riemann},'' \href{http://dx.doi.org/10.1103/PhysRevD.84.044022}{{\em
  Phys.Rev.} {\bf D84} (2011)  044022},
\href{http://arxiv.org/abs/1105.6294}{{\tt arXiv:1105.6294 [hep-th]}}.

N.~Kan, K.~Kobayashi, and K.~Shiraishi, ``{Equations of Motion in Double Field
  Theory: From particles to scale factors},''
  \href{http://dx.doi.org/10.1103/PhysRevD.84.124049}{{\em Phys.Rev.} {\bf D84}
  (2011)  124049},
\href{http://arxiv.org/abs/1108.5795}{{\tt arXiv:1108.5795 [hep-th]}}.

O.~Hohm, S.~K. Kwak, and B.~Zwiebach, ``{Double Field Theory of Type II
  Strings},'' \href{http://dx.doi.org/10.1007/JHEP09(2011)013}{{\em JHEP} {\bf
  1109} (2011)  013},
\href{http://arxiv.org/abs/1107.0008}{{\tt arXiv:1107.0008 [hep-th]}}.

C.~Albertsson, S.-H. Dai, P.-W. Kao, and F.-L. Lin, ``{Double Field Theory for
  Double D-branes},'' \href{http://dx.doi.org/10.1007/JHEP09(2011)025}{{\em
  JHEP} {\bf 1109} (2011)  025},
\href{http://arxiv.org/abs/1107.0876}{{\tt arXiv:1107.0876 [hep-th]}}.

D.~C. Thompson, ``{Duality Invariance: From M-theory to Double Field Theory},''
  \href{http://dx.doi.org/10.1007/JHEP08(2011)125}{{\em JHEP} {\bf 1108} (2011)
   125},
\href{http://arxiv.org/abs/1106.4036}{{\tt arXiv:1106.4036 [hep-th]}}.

I.~Jeon, K.~Lee, and J.-H. Park, ``{Double field formulation of Yang-Mills
  theory},'' \href{http://dx.doi.org/10.1016/j.physletb.2011.05.051}{{\em
  Phys.Lett.} {\bf B701} (2011)  260--264},
\href{http://arxiv.org/abs/1102.0419}{{\tt arXiv:1102.0419 [hep-th]}}.

O.~Hohm and B.~Zwiebach, ``{On the Riemann Tensor in Double Field Theory},''
\href{http://arxiv.org/abs/1112.5296}{{\tt arXiv:1112.5296 [hep-th]}}.

I.~Vaisman, ``{On the geometry of double field theory},''
\href{http://arxiv.org/abs/1203.0836}{{\tt arXiv:1203.0836 [math.DG]}}.

\bibitem{Hohm:2011gs}
O.~Hohm, ``{T-duality versus Gauge Symmetry},''
  \href{http://dx.doi.org/10.1143/PTPS.188.116}{{\em Prog.Theor.Phys.Suppl.}
  {\bf 188} (2011)  116--125},
\href{http://arxiv.org/abs/1101.3484}{{\tt arXiv:1101.3484 [hep-th]}}.

B.~Zwiebach, ``{Double Field Theory, T-Duality, and Courant Brackets},''
\href{http://arxiv.org/abs/1109.1782}{{\tt arXiv:1109.1782 [hep-th]}}.

\bibitem{Aldazabal:2011yz}
G.~Aldazabal, D.~Marques, C.~Nunez, and J.~A. Rosabal, ``{On Type IIB moduli
  stabilization and N = 4, 8 supergravities},''
  \href{http://dx.doi.org/10.1016/j.nuclphysb.2011.03.016}{{\em Nucl.Phys.}
  {\bf B849} (2011)  80--111},
\href{http://arxiv.org/abs/1101.5954}{{\tt arXiv:1101.5954 [hep-th]}}.

\bibitem{Dibitetto:2011eu}
G.~Dibitetto, A.~Guarino, and D.~Roest, ``{How to halve maximal
  supergravity},'' \href{http://dx.doi.org/10.1007/JHEP06(2011)030}{{\em JHEP}
  {\bf 1106} (2011)  030},
\href{http://arxiv.org/abs/1104.3587}{{\tt arXiv:1104.3587 [hep-th]}}.

\bibitem{Dall'Agata:2007sr}
G.~Dall'Agata, N.~Prezas, H.~Samtleben, and M.~Trigiante, ``{Gauged
  Supergravities from Twisted Doubled Tori and Non-Geometric String
  Backgrounds},'' \href{http://dx.doi.org/10.1016/j.nuclphysb.2008.02.020}{{\em
  Nucl.Phys.} {\bf B799} (2008)  80--109},
\href{http://arxiv.org/abs/0712.1026}{{\tt arXiv:0712.1026 [hep-th]}}.

D.~Andriot, R.~Minasian, and M.~Petrini, ``{Flux backgrounds from Twists},''
  \href{http://dx.doi.org/10.1088/1126-6708/2009/12/028}{{\em JHEP} {\bf 0912}
  (2009)  028},
\href{http://arxiv.org/abs/0903.0633}{{\tt arXiv:0903.0633 [hep-th]}}.

\bibitem{Howe:1997qt}
P.~S. Howe, N.~Lambert, and P.~C. West, ``{A New massive type IIA supergravity
  from compactification},''
  \href{http://dx.doi.org/10.1016/S0370-2693(97)01199-4}{{\em Phys.Lett.} {\bf
  B416} (1998)  303--308},
\href{http://arxiv.org/abs/hep-th/9707139}{{\tt arXiv:hep-th/9707139
  [hep-th]}}.

\bibitem{Bergshoeff:2002nv}
E.~Bergshoeff, T.~de~Wit, U.~Gran, R.~Linares, and D.~Roest, ``{(Non)Abelian
  gauged supergravities in nine-dimensions},'' {\em JHEP} {\bf 0210} (2002)
  061,
\href{http://arxiv.org/abs/hep-th/0209205}{{\tt arXiv:hep-th/0209205
  [hep-th]}}.

\bibitem{LeDiffon:2008sh}
A.~Le~Diffon and H.~Samtleben, ``{Supergravities without an Action: Gauging the
  Trombone},'' \href{http://dx.doi.org/10.1016/j.nuclphysb.2008.11.010}{{\em
  Nucl.Phys.} {\bf B811} (2009)  1--35},
\href{http://arxiv.org/abs/0809.5180}{{\tt arXiv:0809.5180 [hep-th]}}.

\bibitem{LeDiffon:2011wt}
A.~Le~Diffon, H.~Samtleben, and M.~Trigiante, ``{N=8 Supergravity with Local
  Scaling Symmetry},'' \href{http://dx.doi.org/10.1007/JHEP04(2011)079}{{\em
  JHEP} {\bf 1104} (2011)  079}, \href{http://arxiv.org/abs/1103.2785}{{\tt
  arXiv:1103.2785 [hep-th]}}.

\bibitem{Romans:1985tz}
L.~Romans, ``{Massive N=2a Supergravity in Ten-Dimensions},''
\href{http://dx.doi.org/10.1016/0370-2693(86)90375-8}{{\em Phys.Lett.} {\bf
  B169} (1986)  374}.

\bibitem{Polchinski:1995mt}
J.~Polchinski, ``{Dirichlet Branes and Ramond-Ramond charges},''
  \href{http://dx.doi.org/10.1103/PhysRevLett.75.4724}{{\em Phys.Rev.Lett.}
  {\bf 75} (1995)  4724--4727},
\href{http://arxiv.org/abs/hep-th/9510017}{{\tt arXiv:hep-th/9510017
  [hep-th]}}.

\bibitem{Hohm:2011cp}
O.~Hohm and S.~K. Kwak, ``{Massive Type II in Double Field Theory},''
  \href{http://dx.doi.org/10.1007/JHEP11(2011)086}{{\em JHEP} {\bf 1111} (2011)
   086},
\href{http://arxiv.org/abs/1108.4937}{{\tt arXiv:1108.4937 [hep-th]}}.

\bibitem{Gates:1984kr}
J.~Gates, S.J., H.~Nishino, and E.~Sezgin, ``{Supergravity in d = 9 and its
  coupling to noncompact sigma model},''
\href{http://dx.doi.org/10.1088/0264-9381/3/1/005}{{\em Class.Quant.Grav.} {\bf
  3} (1986)  21}.

\bibitem{Roest:2004pk}
D.~Roest, ``{M-theory and gauged supergravities},''
  \href{http://dx.doi.org/10.1002/prop.200410192}{{\em Fortsch.Phys.} {\bf 53}
  (2005)  119--230}, \href{http://arxiv.org/abs/hep-th/0408175}{{\tt
  arXiv:hep-th/0408175 [hep-th]}}.

\bibitem{FernandezMelgarejo:2011wx}
J.~Fernandez-Melgarejo, T.~Ortin, and E.~Torrente-Lujan, ``{The general
  gaugings of maximal d=9 supergravity},''
  \href{http://arxiv.org/abs/1106.1760}{{\tt arXiv:1106.1760 [hep-th]}}.

\bibitem{Bergshoeff:2002mb}
E.~Bergshoeff, U.~Gran, and D.~Roest, ``{Type IIB seven-brane solutions from
  nine-dimensional domain walls},''
  \href{http://dx.doi.org/10.1088/0264-9381/19/15/321}{{\em Class.Quant.Grav.}
  {\bf 19} (2002)  4207--4226},
\href{http://arxiv.org/abs/hep-th/0203202}{{\tt arXiv:hep-th/0203202
  [hep-th]}}.

\bibitem{Salam:1984ft}
A.~Salam and E.~Sezgin, ``{d = 8 supergravity},''
  \href{http://dx.doi.org/10.1016/0550-3213(85)90613-3}{{\em Nucl.Phys.} {\bf
  B258} (1985)  284}.

\bibitem{Weidner:2006rp}
M.~Weidner, ``{Gauged Supergravities in Various Spacetime Dimensions},''
  \href{http://dx.doi.org/10.1002/prop.200710390}{{\em Fortsch. Phys.} {\bf 55}
  (2007)  843--945},
\href{http://arxiv.org/abs/hep-th/0702084}{{\tt arXiv:hep-th/0702084}}.

\bibitem{Samtleben:2008pe}
H.~Samtleben, ``{Lectures on Gauged Supergravity and Flux Compactifications},''
  \href{http://dx.doi.org/10.1088/0264-9381/25/21/214002}{{\em
  Class.Quant.Grav.} {\bf 25} (2008)  214002},
  \href{http://arxiv.org/abs/0808.4076}{{\tt arXiv:0808.4076 [hep-th]}}.

\bibitem{Dani:2008}
D.~Puigdom\`enech, ``{Embedding tensor approach to maximal D=8
  supergravity},''. (2008) http://thep.housing.rug.nl/theses.

\bibitem{deRoo:2011fa}
M.~de~Roo, G.~Dibitetto, and Y.~Yin, ``{Critical points of maximal D=8 gauged
  supergravities},'' \href{http://dx.doi.org/10.1007/JHEP01(2012)029}{{\em
  JHEP} {\bf 1201} (2012)  029},
\href{http://arxiv.org/abs/1110.2886}{{\tt arXiv:1110.2886 [hep-th]}}.

\bibitem{GTZ}
B.~T.~P. Gianni and G.~Zacharias, ``{Gr{\"o}bner bases and Primary
  Decomposition of Polynomial Ideals},'' {\em J. Symbolic Computation} {\bf 6}
  (1988)  149--167.

\bibitem{DGPS}
W.~Decker, G.-M. Greuel, G.~Pfister, and H.~Sch{\"o}nemann, ``{\sc Singular}
  {3-1-2} --- {A} computer algebra system for polynomial computations,''.
  http://www.singular.uni-kl.de.

\bibitem{Bergshoeff:2003ri}
E.~Bergshoeff, U.~Gran, R.~Linares, M.~Nielsen, T.~Ortin, and D.~Roest, ``{The
  Bianchi classification of maximal D = 8 gauged supergravities},''
  \href{http://dx.doi.org/10.1088/0264-9381/20/18/310}{{\em Class.Quant.Grav.}
  {\bf 20} (2003)  3997--4014}, \href{http://arxiv.org/abs/hep-th/0306179}{{\tt
  arXiv:hep-th/0306179 [hep-th]}}.

\bibitem{Samtleben:2005bp}
H.~Samtleben and M.~Weidner, ``{The maximal D = 7 supergravities},''
  \href{http://dx.doi.org/10.1016/j.nuclphysb.2005.07.028}{{\em Nucl. Phys.}
  {\bf B725} (2005)  383--419},
\href{http://arxiv.org/abs/hep-th/0506237}{{\tt arXiv:hep-th/0506237}}.

\bibitem{Bergshoeff:2007vb}
E.~A. Bergshoeff, J.~Gomis, T.~A. Nutma, and D.~Roest, ``{Kac-Moody Spectrum of
  (Half-)Maximal Supergravities},''
  \href{http://dx.doi.org/10.1088/1126-6708/2008/02/069}{{\em JHEP} {\bf 0802}
  (2008)  069}, \href{http://arxiv.org/abs/0711.2035}{{\tt arXiv:0711.2035
  [hep-th]}}.

\bibitem{Roest:2009tt}
D.~Roest and J.~Rosseel, ``{De Sitter in Extended Supergravity},''
  \href{http://dx.doi.org/10.1016/j.physletb.2010.01.064}{{\em Phys.Lett.} {\bf
  B685} (2010)  201--207}, \href{http://arxiv.org/abs/0912.4440}{{\tt
  arXiv:0912.4440 [hep-th]}}.

\bibitem{Dibitetto:2010rg}
G.~Dibitetto, R.~Linares, and D.~Roest, ``{Flux Compactifications, Gauge
  Algebras and De Sitter},''
  \href{http://dx.doi.org/10.1016/j.physletb.2010.03.074}{{\em Phys.Lett.} {\bf
  B688} (2010)  96--100}, \href{http://arxiv.org/abs/1001.3982}{{\tt
  arXiv:1001.3982 [hep-th]}}.

\bibitem{Cvetic:2000dm}
M.~Cvetic, H.~Lu, and C.~Pope, ``{Consistent Kaluza-Klein sphere reductions},''
  \href{http://dx.doi.org/10.1103/PhysRevD.62.064028}{{\em Phys.Rev.} {\bf D62}
  (2000)  064028},
\href{http://arxiv.org/abs/hep-th/0003286}{{\tt arXiv:hep-th/0003286
  [hep-th]}}.

\bibitem{Salam:1983fa}
A.~Salam and E.~Sezgin, ``{SO(4) gauging of N=2 supergravity in
  seven-dimensions},''
\href{http://dx.doi.org/10.1016/0370-2693(83)90167-3}{{\em Phys.Lett.} {\bf
  B126} (1983)  295--300}.

\bibitem{deRoo:2006ms}
M.~de~Roo, D.~B. Westra, and S.~Panda, ``{Gauging CSO groups in N=4
  Supergravity},'' \href{http://dx.doi.org/10.1088/1126-6708/2006/09/011}{{\em
  JHEP} {\bf 0609} (2006)  011},
  \href{http://arxiv.org/abs/hep-th/0606282}{{\tt arXiv:hep-th/0606282
  [hep-th]}}.

\end{thebibliography}\endgroup
\bibliographystyle{utphys}

\end{document}